\documentclass[12pt]{article}
\usepackage{fullpage,graphicx,psfrag,url,ar,rotating,url,hyperref}
\usepackage[small,bf]{caption}
\usepackage{amsmath,amssymb,enumitem,bbm,subcaption,multirow}
\newcommand{\BEAS}{\begin{eqnarray*}}
\newcommand{\EEAS}{\end{eqnarray*}}
\newcommand{\BEQ}{\begin{equation}}
\newcommand{\EEQ}{\end{equation}}
\newcommand{\BIT}{\begin{itemize}}
\newcommand{\EIT}{\end{itemize}}
\newcommand{\BMAT}{\begin{bmatrix}}
\newcommand{\EMAT}{\end{bmatrix}}

\newcommand{\eg}{{\it e.g.}}
\newcommand{\ie}{{\it i.e.}}

\newcommand{\reals}{{\mbox{\bf R}}}
\newcommand{\integers}{{\mbox{\bf Z}}}
\newcounter{oursection}

\newcommand{\conrad}{\texttt{ConRad}}

\newcounter{algorithmctr}[section]
\renewcommand{\thealgorithmctr}{\thesection.\arabic{algorithmctr}}
\newenvironment{algdesc}%                                                       
   {\refstepcounter{algorithmctr}\begin{list}{}{%                               
       \setlength{\rightmargin}{0\linewidth}%                                   
       \setlength{\leftmargin}{.05\linewidth}}%                                 
       \rmfamily\small
       \item[]{\setlength{\parskip}{0ex}\hrulefill\par%                         
        \nopagebreak{\bfseries\textsf{Algorithm \thealgorithmctr~}}}}%          
   {{\setlength{\parskip}{-1ex}\nopagebreak\par\hrulefill} \end{list}}

\title{\Large{
A Convex Optimization Approach to 
Radiation Treatment Planning with Dose Constraints}}
\author{Anqi Fu\thanks{Anqi Fu and Bar\i\d s Ungun contributed equally to this paper.} \and Bar\i\d s Ungun\footnotemark[1] \and Lei Xing \and Stephen Boyd}

\begin{document}
\maketitle

\begin{abstract}
We present a method for handling dose constraints as part of a convex 
programming framework for inverse treatment planning. Our method uniformly handles mean dose, maximum dose, minimum dose, and dose-volume (\ie, percentile) constraints as part of a convex formulation. 
Since dose-volume constraints are non-convex, we replace them with a convex 
restriction. 
This restriction is, by definition, conservative; to mitigate its 
impact on the clinical objectives, we develop a two-pass planning algorithm 
that allows each dose-volume constraint to be met exactly on a second pass by the solver if its corresponding restriction is feasible on the first pass. 
In another variant, we add slack variables to each dose constraint to 
prevent the problem from becoming infeasible when the user specifies an 
incompatible set of constraints or when the constraints are made 
infeasible by our restriction.
Finally, we introduce \conrad{}, a Python-embedded open-source software package for 
convex radiation treatment planning. \conrad{} implements the methods described 
above and allows users to construct and plan cases through a simple interface.
\end{abstract}
%\tableofcontents
%\newpage

\section{Introduction}\label{intro}

External beam radiation therapy is the treatment of diseased tissue with beams 
of ionizing radiation delivered from a source outside the patient. 
When radiation passes through the patient, it damages both healthy and diseased 
tissue.
A treatment plan must be carefully designed to minimize harm to healthy organs, 
while delivering enough dose to ablate the targeted tissue. 
With recent hardware advances, delivered beams can be positioned and shaped with growing sophistication, and clinicians rely increasingly on optimization techniques to 
guide their treatment decisions. 
In this paper, we focus on one part of the treatment planning process: 
the selection of an optimal intensity profile for every radiation beam.

In \cite{SFOMackie:1999}, the authors provide a comprehensive survey of several problem 
formulations for treatment planning, including linear \cite{RLMBelli:1991} 
\cite{Holder:2003} and quadratic programming models \cite{BBBSchlegel:1990, 
XingChen:1996, XHS+Boyer:1998}. 
Generally, linear models minimize the weighted sum of doses or the maximum 
deviation from a prescribed dose, while quadratic models minimize the weighted 
sum of squared difference between actual and prescribed dose. 
These formulations incorporate linear bounds on the dose to each structure. Solutions can be rapidly found using various interior point methods, such as primal-dual \cite{AlemanGlaser:2010}, projected gradient \cite{AlemanMisic:2013}, and interior point constraint generation \cite{Oskoorouchi:2011}.

To address conflicting clinical goals, researchers have proposed models with multiple objectives and 
constraints. By varying the weight on each objective, one can produce a set of solutions on 
the Pareto frontier \cite{HamacherKufer:2002} \cite{HCBortfeld:2006}. A large number of treatment evaluation criteria can be transformed into convex criteria within this framework \cite{RomeijnDempseyLi:2004}. Although multi-objective optimization offers flexibility, calculating thousands of points on the 
Pareto frontier proves computationally inefficient in practice, and expert judgment is still 
required to select a clinically acceptable plan from the set of mathematically 
optimal plans.

All the methods discussed so far hinge on a convex problem formulation. 
However, many clinically relevant constraints are non-convex. 
One such type is the dose-volume constraint, which bounds the dose delivered to 
a given percentage of a patient's anatomy \cite{ZLL+Jiang:2014}. 
A review of some models for handling this class of constraints is given in \cite{EGHShao:2008}. 
The simplest approach is to add a nonlinear, volume-sensitive penalty to the 
objective function \cite{CLM+Phillips:1998} \cite{SpirouChui:1998}. 
Then, a local search algorithm, such as the conjugate gradient method 
\cite{XingChen:1996} \cite{XHS+Boyer:1998} \cite{SORMackie:2000} or simulated 
annealing \cite{Webb:1989} \cite{Webb:1992} \cite{MagerasMohan:1993}, is used to 
solve the optimization problem. 
Unfortunately, since this formulation is non-convex, these algorithms often 
produce a local minimum, resulting in a sub-optimal treatment plan 
\cite{Deasy:1997} \cite{WuMohan:2002}. 
Another method is to directly model the dose-volume constraint with a set of 
binary decision variables. 
Each variable indicates whether a voxel should be included in the fraction of 
a structure's volume that must fulfill the dose bound \cite{LBU+Shapiro:1990} 
\cite{LFCrocker:2000} \cite{LFCrocker:2003}. 
Given the dimensions of patient data, this results in a large-scale 
mixed-integer programming problem, which is prohibitively expensive to compute 
for most clinical cases.

A more promising approach is to replace each dose-volume constraint with a 
convex approximation. 
This allows users to take advantage of large-scale convex optimization algorithms 
to quickly generate treatment plans. For instance, \cite{HalabiCraft:2006} substitutes a ramp function for the indicator that a particular voxel violates its desired dose-volume threshold, then penalizes the total number of voxel violations in the objective. Other researchers have employed the conditional value-at-risk (CVaR), a metric that represents the average tail loss in a probability distribution \cite{RockafellarUryasev:2000}. 
It is convex in the loss variable and thus offers a computationally suitable 
alternative to the dose-volume constraint. In the recent literature, CVaR has been used to formulate linear constraints on the 
average dose in the upper and lower tails of a structure's dose distribution, 
leading to significant improvements in treatment plans 
\cite{RAD+Li:2003} \cite{RADKumar:2006} \cite{CMPurdie:2014}. 
However, CVaR functions are parametric, and implementations of this model require 
a heuristic search over the parameter space to obtain a good approximation of 
the dose-volume constraint \cite{AGS+Schreibmann:2010}.

Perhaps the method most similar to ours is \cite{ZST+Zinchenko:2013}. In this paper, the authors propose constraining the dose moments to equal those of the desired dose-volume histogram curve. They derive a convex relaxation of these constraints, then solve their treatment planning problem in two phases: the first phase adds slack variables to the moment bounds, so a solution is always feasible, while the second phase tightens these bounds in order to improve upon plan quality whenever possible. Using only three moments, their technique is able to closely match the reference histogram curves in a prostate cancer case.

These results demonstrate the power of convex models: they provide more 
flexibility than linear or quadratic models, while avoiding the issue of 
multiple local optima in non-convex formulations and the intractability of a 
mixed-integer program. 
In this paper, we propose a new convex formulation of the fluence map 
optimization problem with dose-volume constraints. 
Given a predetermined number of candidate beams, we construct a convex 
optimization problem around a set of clinical goals and solve for the radiation 
intensity pattern that best achieves these goals subject to restrictions on the 
dose distribution. 
Our algorithm is quick and efficient, allowing clinicians to rapidly compare 
trade-offs between different plans and select the best treatment for a patient. We provide a Python package, \href{https://github.com/bungun/conrad}{\conrad{}}, that implements our method within a simple intuitive interface.

This paper is organized as follows: 
In Sect. \ref{physics}, we review the radiation treatment planning problem. 
In Sect. \ref{planning}, we describe the patient characteristics and constraints that 
clinicians must consider when selecting an optimal plan. 
In Sect. \ref{objective}, we define a convex optimization framework for the 
basic treatment planning problem. 
Sect. \ref{constraints} introduces dose constraints. 
We show how to incorporate dose-volume constraints via a convex restriction, 
which provides an approximation of the dose percentile. 
In Sect. \ref{refinements}, we present two extensions to our model. 
Sect. \ref{conrad} describes the Python implementation of our algorithm, 
and Sect. \ref{examples} demonstrates its performance in several clinical 
cases. 
Finally, Sect. \ref{conclusion} concludes.

\section{Problem Description}\label{physics}

During external beam radiation therapy, ionizing radiation travels through a 
patient, depositing energy along the beam paths. 
Radiation damages both diseased and healthy tissue, but clinicians aim to damage
these tissues differentially, exploiting the fact that cancer cells typically have faulty cell repair mechanisms and exhibit a lower tolerance to radiation than healthy cells. 
The goal is to focus radiation 
beams such that enough dose is delivered to kill diseased tissue, while avoiding as much of the surrounding healthy organs as possible. 
The clinician separates these structures into one or more planning target volumes (PTVs) to be irradiated at a prescribed dose level and several organs-at-risk (OARs) to spare from radiation. 

Before treatment, the patient is positioned on a couch. Photon, electron, 
proton, or heavier particle beams are generated with a particle accelerator and 
coupled to a mechanized gantry that contains additional hardware components, which
shape and focus the beams. The gantry typically rotates around one central axis
(but may have additional rotational and translational degrees of freedom
\cite{Mackie:1993} \cite{GlideHurst:2013} \cite{Adler:1998}), so that by controlling the gantry 
and couch, beams can be delivered from almost any angle and location around the 
patient. 

Delivery strategies vary from using a large number of apertures (beam
shapes) delivered sequentially from a few beam angles, as in 
intensity-modulated radiation therapy (IMRT), to calculating a single optimal 
aperture at a large number of angles, as in volumetric modulated arc therapy 
(VMAT). 
For a given delivery strategy, the goal of treatment planning is to determine the optimal beam angles, shapes and intensities that most closely approximate a 
desired dose distribution to the targeted volumes. 
In this work, we consider the task of optimizing intensities for a given set of 
beams of known positions and shapes, \ie, calculating optimal beam weights. This is applicable to the fluence map optimization (FMO) step in IMRT planning, 
the FMO step in direct aperture optimization for modalities such as VMAT, 4$\pi$, or SPORT \cite{Bedford:2009} \cite{Dong:2013} \cite{LiXing:2013}, as well as inverse planning problems for other common modalities such as stereotactic radiosurgery \cite{ShepardFerris:2000} \cite{SchweikardSchlaefer:2006} or proton beam therapy \cite{OelfkeBortfeld:2001}.

\section{Clinical Planning}\label{planning}

\subsection{Dose Physics}

Prior to treatment, medical images---such as CT, MRI and PET scans---are 
collected to form a three-dimensional image of the patient's anatomy. 
This representation is discretized into regular volume elements, or voxels. 
The anatomy is then delineated by clinicians into various structures, and the 
dose to each structure is considered during planning.
Although the structure contours drawn by clinicians may overlap, in this work, we
associate each voxel with a single structure. 

Dose calculation algorithms range from analytical approximations to Monte Carlo
simulations, but in all cases, they provide a model with a linear mapping from 
beam intensities to delivered voxel doses. 
The dose within each voxel is assumed to be uniform.
For each candidate beam, we have an aperture shape that may
be further subdivided, \eg, into regular rectangular subdivisions called
beamlets.
The intensities of these beams (or beamlets) are represented in a vector. 
A patient-specific dose deposition matrix maps this vector of radiation 
intensities to the vector of doses delivered per voxel.

\subsection{Dose Objectives}

Given a fixed number of candidate beams, our goal is to determine the beam 
intensities that satisfy a clinical objective defined in terms of the dose 
delivered to each voxel in the patient anatomy.

Every PTV is prescribed a desired dose, which we wish to deposit 
uniformly throughout the target. 
Delivering too high or too low a dose of radiation has different clinical 
consequences, so we introduce separate underdose and overdose penalties 
for every PTV. 
In the case of OARs, a lower dose is always preferable, so we penalize any dose
above zero and omit an underdose penalty term. 
In addition to clinical considerations, such as the patient's medical history and
past courses of radiation therapy, different organs usually exhibit different levels of sensitivity to radiation.
For these reasons, we allow the penalties for each OAR to be scaled 
independently, allowing the planner to adjust the relative importance of meeting 
the dose targets for each structure separately.

We apply the penalty associated with each structure to every voxel in that
structure, and the objective function of our treatment planning problem tallies 
these dose penalties over all voxels in a patient's anatomy.

\subsection{Dose Constraints}

In addition to dose penalties, we allow for hard constraints on the amount of 
radiation delivered to portions of the patient anatomy. 
For example, the clinician may only consider plans in which the spinal column 
receives a dose below a certain level because any more will increase the likelihood of injury beyond an acceptable limit. 
Basic constraints of this nature take the form of bounds on the mean, minimum, 
and maximum dose to a structure. 
More generally, bounds can be enforced on the dose to a fraction of the voxels 
in a structure. 
These dose-volume constraints restrict the relative volume that receives 
radiation beyond a particular threshold, giving the clinician precise control 
over the dose distribution. 
This is especially important when sparing OARs, since some organs are able to 
sustain high levels of uniform radiation, while others will fail unless the 
radiation is contained to a small fraction of the tissue.

Clinicians typically use a dose-volume histogram (DVH) to assess the quality of 
a treatment plan. 
For every structure, the DVH specifies the percentage of its volume that 
receives at least a certain dose. 
A point $(x,y)$ on the curve indicates that $y\%$ of the total voxels in the 
structure receives a dose of at least $x$ Gy. 
Ideally, we want our structures to receive exactly the prescribed dose 
throughout their volumes. 
If the prescription is $d$ Gy, then our optimal DVH curve for the PTV is a step 
function with a drop at $(d,100)$, and our optimal DVH for each OAR exhibits a 
drop at $(0,100)$.

\begin{figure}
	\begin{center}
		\begin{subfigure}[b]{0.48\textwidth}
			\caption{Lower Dose-Volume Constraint}
			\includegraphics[width=\textwidth]{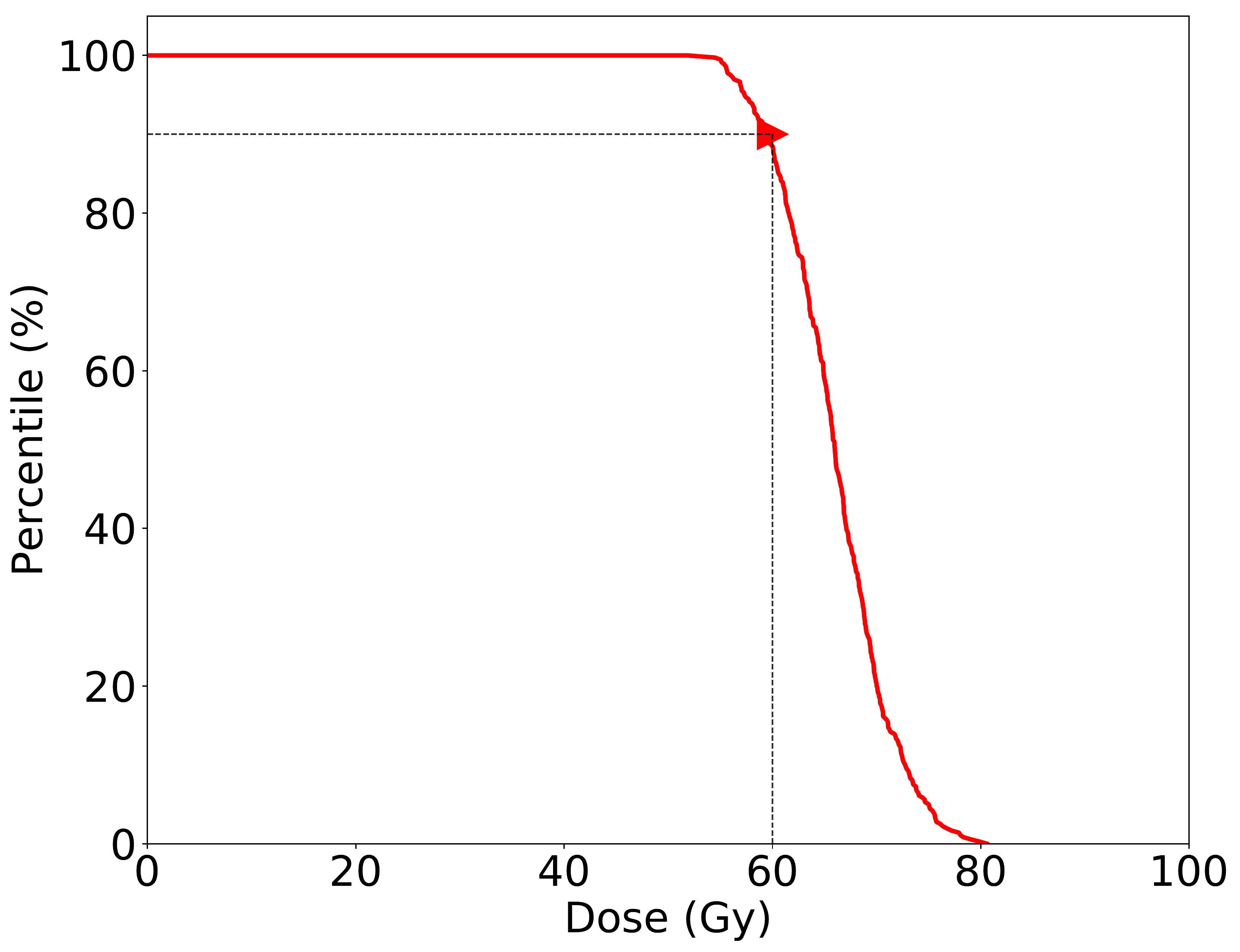}
		\end{subfigure}
		\begin{subfigure}[b]{0.48\textwidth}
			\caption{Upper Dose-Volume Constraint}
			\includegraphics[width=\textwidth]{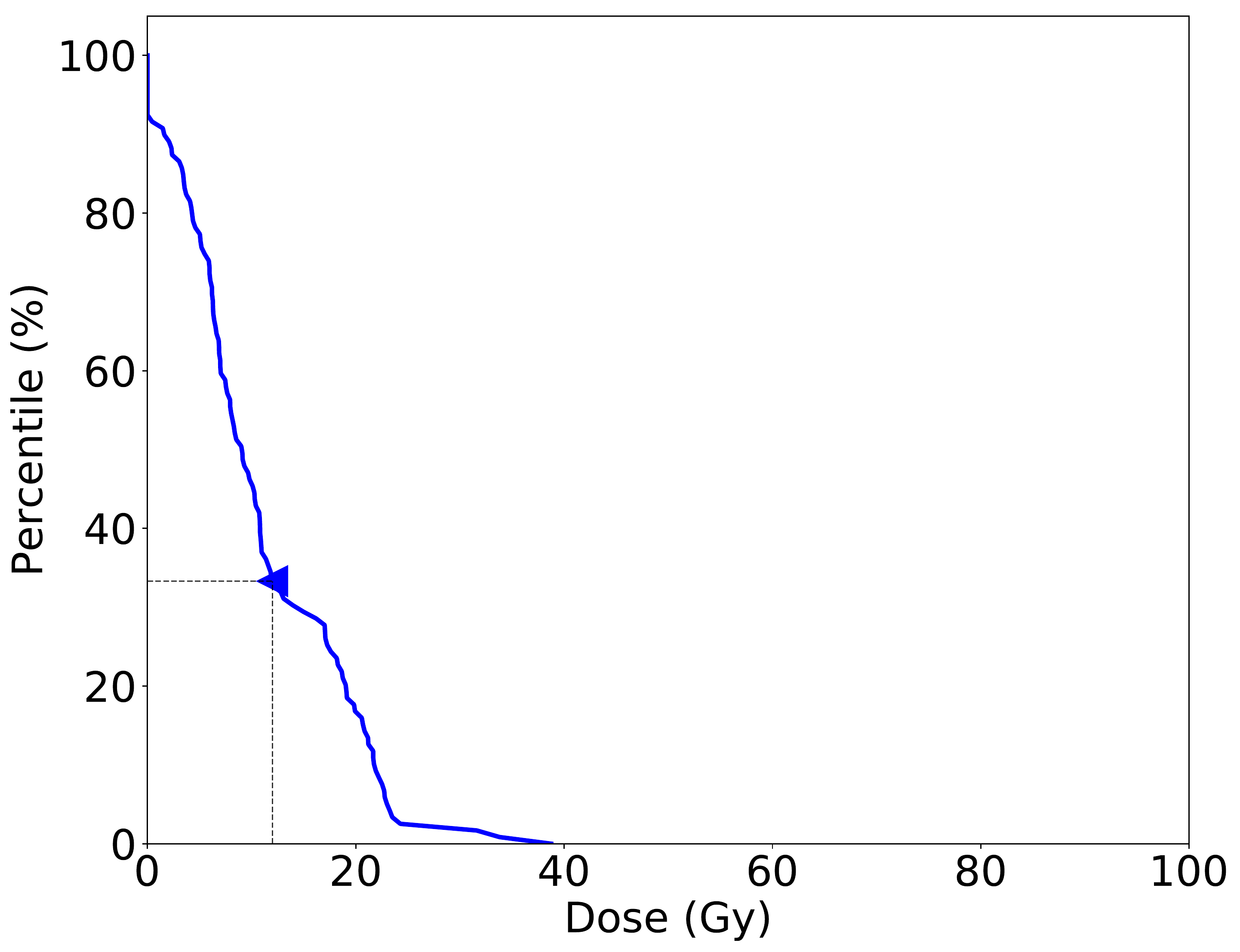}
		\end{subfigure}
	\end{center}
	\caption{(a) A lower DVH constraint ensures at least 90\% of the structure's volume receives at least 60 Gy. The dotted line intersects the curve at $(60,90)$. (b) An upper DVH constraint allows at most 33\% of the volume to receive at least 12 Gy.}
	\label{f-dvh-con}
\end{figure}

Dose constraints restrict the shape and location of points on the DVH curve. In Fig.~\ref{f-dvh-con}, a lower dose-volume constraint, $D(90) \geq 60$, is represented by the right-facing arrow centered at $(60, 90)$. This ensures that a minimum of 90\% of the structure's volume receives at least 60 Gy, \ie, $y \geq 90$ along the vertical line $x = 60$. The PTV's DVH curve is pushed rightward by this type of constraint. Similarly, an upper dose-volume constraint, $D(33) \leq 12$, is labeled with a left-facing arrow at $(12,33)$, which pushes the OAR's DVH curve leftward, representing the restriction that $y \leq 33$ along the line $x = 12$. Together, the DVH curves and their respective dose constraints enable the clinician to easily visualize trade-offs when formulating a treatment plan.

\section{Convex Formulation}\label{objective}

Consider a case with $m$ voxels inside a patient volume and $n$ candidate 
treatment beams. 
Our goal is to determine the beam intensities $x \in \reals_+^n$ that deliver a 
vector of voxel doses $y \in \reals_+^m$, which meet a set of clinical 
objectives. We are given a case-specific dose influence matrix $A \in \reals_+^{m \times n}$ 
that approximates the relationship between the beams and doses linearly as 
$y = Ax$. We refer to the rows of $A$ as $a_i \in \reals_+^n$ for $i = 1,\ldots,m$.
The basic inverse treatment planning problem is of the form
\[
\begin{array}{ll}
	\mbox{minimize}  & f(y) \\
	\mbox{subject to} 
		& y=Ax \\
		& x \succeq 0
\end{array}
\]
with respect to $x$ and $y$, where $f: \reals^m \rightarrow \reals$ is a convex loss function chosen to 
penalize voxel doses based on the goals of the clinician. 
Here, the inequality on $x$ is understood to be applied element-wise. 
In a typical case, a patient is prescribed a treatment plan that can be 
characterized by a vector of doses $d \in \reals_+^m$ to each voxel. 
Our function $f$ then penalizes the deviation of the calculated dose $y$ from 
the prescribed dose $d$, taking into account the different structures inside a 
patient.

In our formulation, we consider a loss function $f(y) = \sum_{i=1}^m f_i(y_i)$ 
where $y_i = a_i^Tx$ and each $f_i$ is a piecewise-linear function
\[
f_i(y_i) = w_i^-(y_i - d_i)_- + w_i^+(y_i - d_i)_+.
\]
The parameters $w_i^-$ and $w_i^+$ are the non-negative weights on the 
underdose and overdose, respectively. This penalty structure is common in the literature \cite{LimCao:2012} \cite{ChenUnkelbach:2012} and provided the most efficient software implementation.

Prior to treatment planning, the $m$ voxels in a patient volume are grouped 
into $S$ distinct, non-overlapping sets representing the planning target volume 
(PTV), organs-at-risk (OARs), and generic non-target tissue (often labeled 
``body''). 
Each set $\mathcal{V}_s$ contains all the voxel indices $i$ within a 
corresponding internal structure with index $s$. 
Together, $\{\mathcal{V}_s\}_1^S$ forms a partition of the patient volume, \ie, 
$\bigcup_1^S \mathcal{V}_s$ covers all voxel indices and 
$\mathcal{V}_{s_1} \bigcap \mathcal{V}_{s_2} = \emptyset$ for $s_1 \neq s_2$. 
We assume the indices are ordered such that 
$s=1,\ldots,P \leq S$ are targets and the rest non-targets.

\begin{figure}
	\begin{center}
		\begin{subfigure}[b]{0.48\textwidth}
			\caption{PTV Loss Function}
			\includegraphics[width=\textwidth]{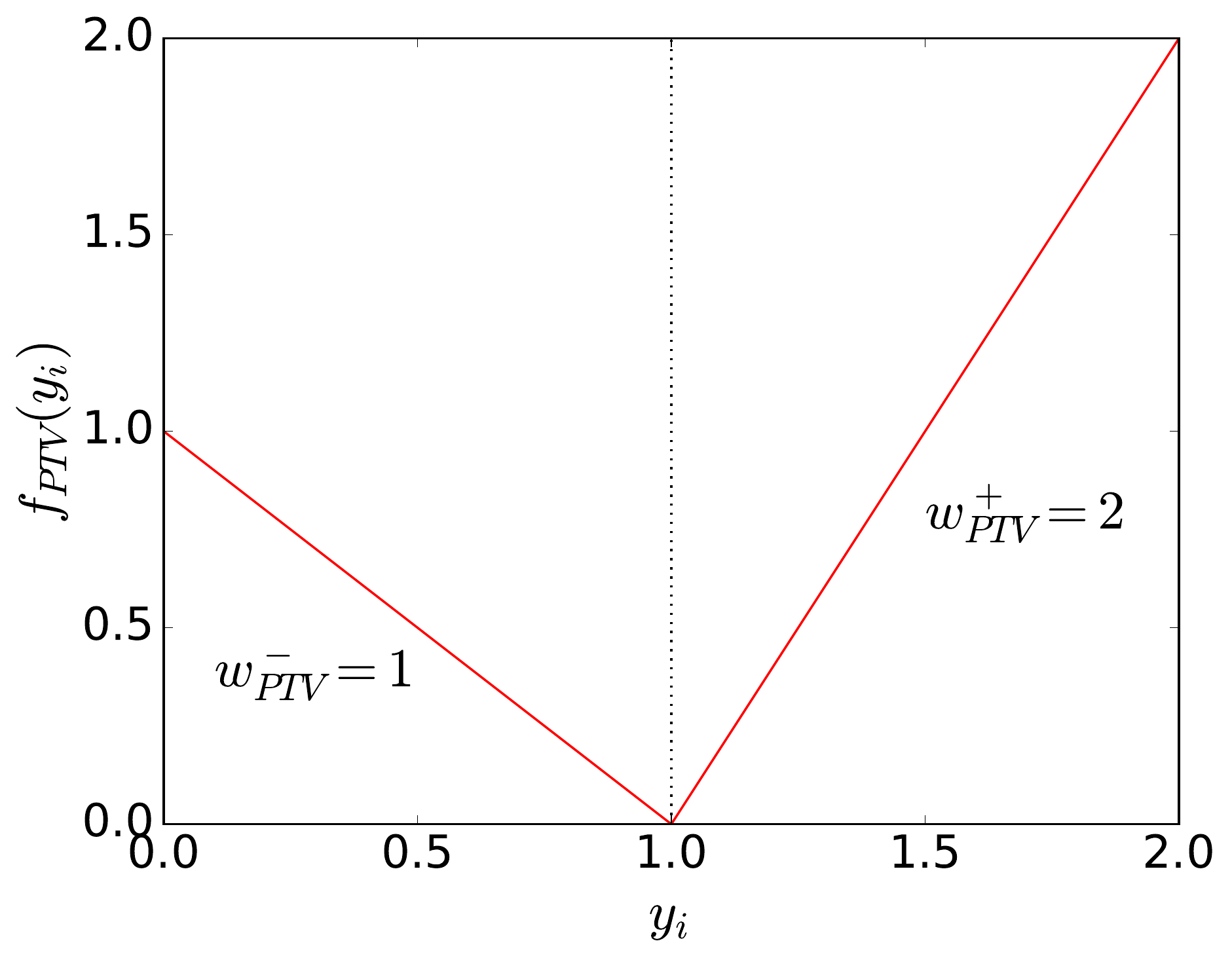}
		\end{subfigure}
		\begin{subfigure}[b]{0.48\textwidth}
			\caption{OAR Loss Function}
			\includegraphics[width=\textwidth]{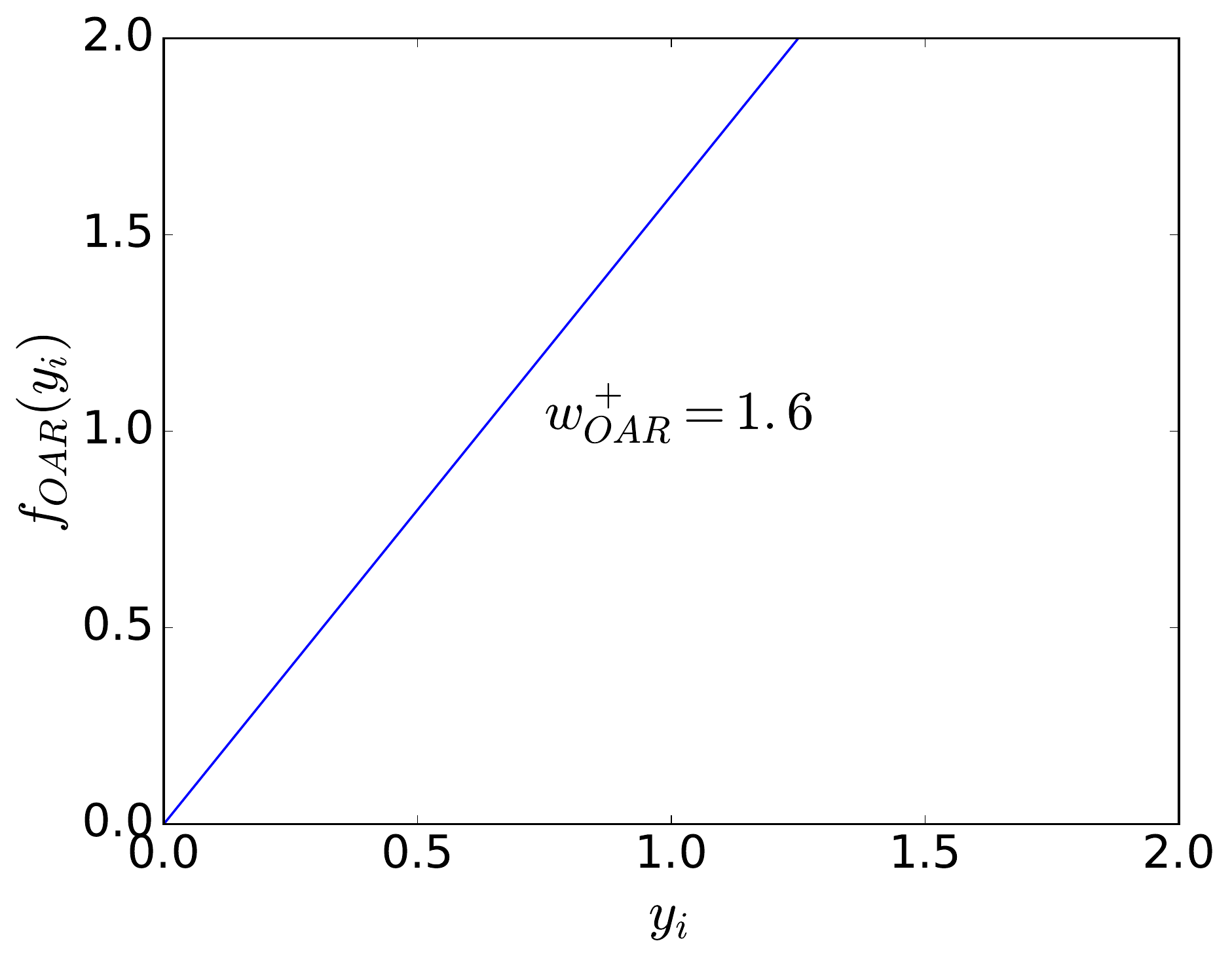}
		\end{subfigure}
	\end{center}
	\caption{(a) The loss function for a PTV prescribed $d_s = 1$ with penalties 
	$w_s^- = 1$ and $w_s^+ = 2$, and (b) the loss function for an OAR with 
	penalty $w_{s'}^+ = 1.6$.}
	\label{f-loss-funcs}
\end{figure}

For simplicity, we choose our voxel doses and penalties to be uniform within 
each structure. 
We let $d_s$ represent the prescribed dose, and $w_s^-$ and $w_s^+$ the 
underdose and overdose penalties for all voxels $i \in \mathcal{V}_s$. 
The loss function for our basic inverse planning problem is 
$f(y) = \sum_{s=1}^S f_s(y_i)$ where
\[
f_s(y_i) 
= \sum_{i \in \mathcal{V}_s} f_i(y_i) 
= \sum_{i \in \mathcal{V}_s} \{ w_s^-(y_i - d_s)_- + w_s^+(y_i - d_s)_+ \}.
\]
A non-target structure $s$ is always prescribed a dose of zero, and since 
$y \geq 0$, its individual loss simplifies to $f_i(y_i) = w_s^+y_i$. 
Thus, its only contribution to the objective is through its total dose. 
An example of these loss functions is given in Fig. \ref{f-loss-funcs}. 
We can collapse the sum of non-target losses into a single linear term,
\[
\sum_{s=P+1}^S f_s(y_i) 
= \sum_{s=P+1}^S w_s^+ \left(\sum_{i \in \mathcal{V}_s} y_i\right) 
= \sum_{s=P+1}^S w_s^+ z_s = c^Tz,
\]
where $c = (w_{P+1}^+,\ldots,w_S^+)$ and 
$z = \left(
\sum_{i \in \mathcal{V}_{P+1}} y_i,\ldots,
\sum_{i \in \mathcal{V}_S} y_i
\right)$. 
Our objective is then
\[
f(y) = \sum_{s=1}^P \sum_{i \in \mathcal{V}_s} \{ 
		w_s^-(y_i - d_s)_- + w_s^+(y_i - d_s)_+ \} + c^Tz.
\]

This formulation is closely related to quantile regression \cite{DFVistocco:2013}. 
In the latter, we minimize $\sum_i \phi(y_i-v-d_i)$ with respect to $(x,v)$ 
where
\[
\phi(u) 
= \tau (u)_+ + (1-\tau)(u)_- 
= \frac{1}{2}|u| + \left(\tau-\frac{1}{2}\right)u
\]
is the tilted $\ell_1$ penalty with $\tau \in (0,1)$. 
For our inverse planning problem, the residual $r_i := y_i-v-d_i$ can be 
interpreted as the difference between calculated and desired doses, allowing for 
a uniform dose offset $v \in \reals$ within each structure. 
We rewrite our individual loss as
\[
\begin{array}{ll}
f_i(y_i-v)
&= w_i^-(r_i)_- + w_i^+(r_i)_+ \\
&= (w_i^- + w_i^+)\left(
		\frac{w_i^-}{w_i^- + w_i^+} (r_i)_- + 
		\frac{w_i^+}{w_i^- + w_i^+} (r_i)_+
	\right) \\
&= (w_i^- + w_i^+)\left(
		\frac{1}{2}|r_i| + 
		\left(\frac{w_i^+}{w_i^-+w_i^+} - \frac{1}{2}\right)r_i
	\right),
\end{array}
\]
and the loss function becomes
\[
f_s(y_i-v) 
= \sum_{i \in \mathcal{V}_s} f_i(y_i-v)
= (w_s^- + w_s^+) \sum_{i \in \mathcal{V}_s} \left(
		\frac{1}{2}|r_i| + \left(\tau_s-\frac{1}{2}\right)r_i
	\right),
\]
where $\tau_s := \frac{w_s^+}{w_s^-+w_s^+} \in (0,1)$. 
For $r_i \neq 0$, the first order condition with respect to $v$ is
\[
\frac{\partial f_s(y_i-v)}{\partial v} 
= \tau_s|\{i: r_i > 0\}| - (1-\tau_s)|\{i: r_i < 0\}| 
= 0,
\]
which implies $\tau_s|\mathcal{V}_s| = |\{i: r_i < 0\}|$, \ie, in a given 
structure $s$, the $\tau_s$-quantile of optimal residuals is zero. 
Although our original loss does not include $v$, we can use this as a rule of 
thumb for selecting relative dose penalties $(w_s^-,w_s^+)$.

\section{Dose Constraints}\label{constraints}

\subsection{Percentile}

A percentile constraint, otherwise known as a dose-volume constraint, bounds the 
dose delivered to a given percentile of a patient structure. This allows us to 
set a limit on the fraction of total voxels that are under- or overdosed with 
respect to a user-provided threshold. 
For clinicians, this provides a way to shape the dose-volume histogram directly 
rather than by searching through combinations of objective weights to achieve 
desired dose statistics. 
Given a structure $s$ and dose vector $y$, let $D_s(p,y)$ represent the minimum
dose delivered to $p$ percent of all voxels in $s$, \ie, $D_s(p,y)$ is the 
greatest lower bound on the dose received by $p\%$ of the tissue. 

To formalize this notion, we define an exact value count function
$v_s: \reals_+^m \times \reals_+ \rightarrow \integers_+$, which computes the 
total number of voxels $i \in \mathcal{V}_s$ that receive a dose above 
$b \in \reals_+$. Let $g(u) = \mathbbm{1}\{ u \geq 0\}$, then
\[
v_s(y,b) 
= \sum_{i \in \mathcal{V}_s} \mathbbm{1}\{ y_i \geq b \} 
= \sum_{i \in \mathcal{V}_s} g(y_i - b)
\]
and our $p$-th percentile dose is
\[
D_s(p,y) = \max \{ b \in \reals_+: v_s(y,b) \geq \phi_s(p) \} 
\quad \mbox{where} \quad \phi_s(p) := \frac{p}{100}|\mathcal{V}_s|.
\]
Observe that $D_s(p,y) \geq 0$ is finite and weakly decreasing in $p$.

Our goal is to bound $D_s(p,y)$. For example, we may want at least $30\%$ of the 
voxels in structure $s$ to receive a dose above 25 Gy; this is identical to 
$D_s(30,y) \geq 25$. 
Let $\ell < u$ be non-negative scalar values. 
An lower dose-volume constraint, $D_s(p,y) \geq \ell$, requires the number of 
voxels in $s$ that receive a dose above $\ell$ to be at least $p\%$ of the total 
voxels in the structure.
Similarly, an upper dose-volume constraint, $D_s(p,y) \leq u$, requires the 
number of voxels $i \in \mathcal{V}_s$ with a dose above $u$ to be at most $p\%$ 
of voxels in $\mathcal{V}_s$, or equivalently, at least $100-p\%$ of the voxels
to receive a dose under $u$. Thus, the inequalities
\[
D_s(p, y) \leq u 
\quad \Leftrightarrow \quad v_s(y, u) \leq \phi_s(p) 
\quad \Leftrightarrow \quad v_s(-y, -u) \geq \phi_s(100-p)
\]
are equivalent, as are
\[
D_s(p, y) \geq \ell 
\quad \Leftrightarrow \quad v_s(y, \ell) \geq \phi_s(p) 
\quad \Leftrightarrow \quad v_s(-y, -\ell) \leq \phi_s(100-p).
\]

In general, this is a hard combinatorial problem: the brute force approach for 
a single upper dose-volume constraint, for example, is to solve all 
$\binom{|\mathcal{V}_s|}{\phi}$ convex problems obtained by choosing subsets of 
$\phi = \lceil{\phi_s(p)}\rceil$ voxels to constrain below $u$, which is 
prohibitively large given the size of patient geometries.

\subsection{Mean, Minimum, and Maximum}

In a few special cases, we can set convex constraints on the dose. 
Let the average, minimum, and maximum dose delivered to all voxels in structure 
$s$ be
\[
D_s^{\mathrm{avg}}(y) = \frac{1}{|\mathcal{V}_s|} \sum_{i \in \mathcal{V}_s} y_i, 
\quad D_s^{\min}(y) = \underset{i \in \mathcal{V}_s}{\min} \{ y_i \},
\quad D_s^{\max}(y) = \underset{i \in \mathcal{V}_s}{\max} \{ y_i \}.
\]
A lower bound $b \in \reals_+$ on the minimum dose is equivalent to requiring 
$y_i \geq b$ for all $i \in \mathcal{V}_s$, and similarly for an upper bound on 
the maximum dose. 
Thus, we can enforce linear constraints on these dose statistics in our problem. 
Our non-convex formulation with exact dose-volume constraints is
\BEQ\label{cvx-exact}
\begin{array}{lll}
	\mbox{minimize} & f(y) & \\
	\mbox{subject to} 
		& y=Ax & \\
		& x \succeq 0 & \\
		& \ell_{s,k} \leq D_s(p_{s,k},y) \leq u_{s,k}, 
				& k=1,\ldots,K_s, \quad s=1,\ldots,S \\
		& \ell_s^{\mathrm{avg}} \leq D_s^{\mathrm{avg}}(y) \leq u_s^{\mathrm{avg}}, 
				& s=1,\ldots,S \\
		& D_s^{\min}(y) \geq \ell_s^{\min}, & s=1,\ldots,S \\
		& D_s^{\max}(y) \leq u_s^{\max}, & s=1,\ldots,S,
\end{array}
\EEQ
where $(x,y)$ are our variables, and for each structure $s$, we index the parameters of its dose-volume 
constraints with $k=1,\ldots,K_s$.

\subsection{Convex Restriction}
% From ConRad package comments:
% upper constraint:
% 	\sum (beta + (Ax - dose + slack)))_+ <= beta * voxel_limit
% lower constraint:
% 	\sum (beta - (Ax - dose - slack)))_+ <= beta * voxel_limit

\begin{figure}
	\begin{center}
		\includegraphics[height=.4\textwidth]{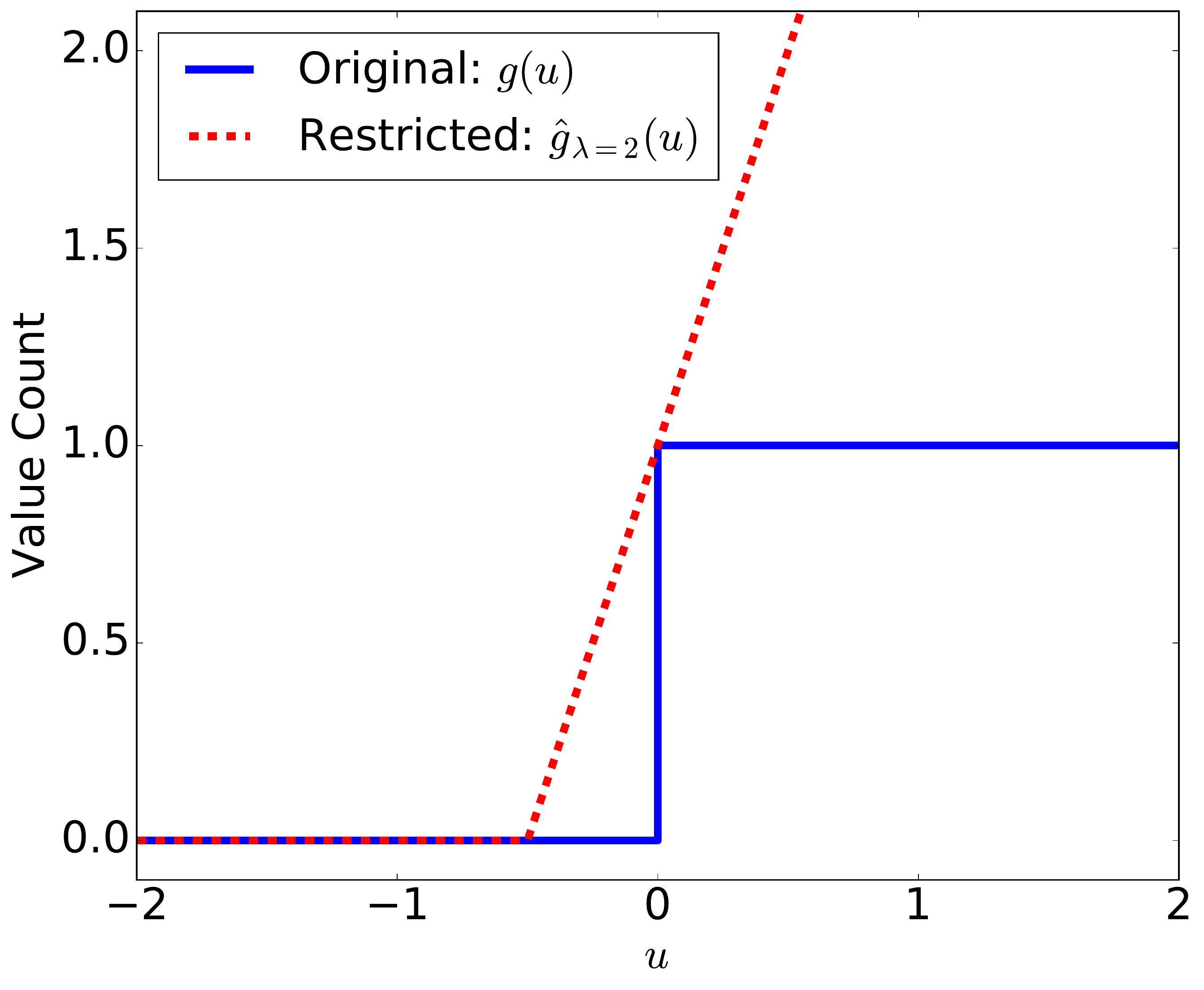}
	\end{center}
	\caption{The indicator function $g(u)$ (solid) and hinge loss 
	$\hat g_{\lambda}(u)$ with $\lambda = 2$ (dashed). Note that 
	$\hat g_{\lambda}(u) \geq g(u)$ for all $u \in \reals$, so the hinge loss 
	provides a convex restriction on the dose-volume constraint.}
	\label{f-hinge-restrict}
\end{figure}

To address the non-convexities in Prob. \ref{cvx-exact}, we introduce a convex 
restriction that provides an effective heuristic for satisfying the dose-volume 
constraints. 
Our restricted constraint overestimates the number of voxels that are underdosed 
with respect to $d$ by replacing the indicator $g$ in $v$ with a family of hinge 
loss functions 
\[
\hat g_{\lambda}(u) = (1+\lambda u)_+ = \max(1+\lambda u,0),
\]
parametrized by $\lambda > 0$, giving us a restricted value count for structure $s$ of
\[
\hat v_s(y,b;\lambda) 
= \sum_{i \in \mathcal{V}_s} \hat g_{\lambda}(y_i - b) 
= \sum_{i \in \mathcal{V}_s} (1 + \lambda(y_i - b))_+.
\]
If $u > 0$, then $g(u) = 1 < 1+\lambda u = \hat g_{\lambda}(u)$, and if 
$u \leq 0$, then $g(u) = 0 \leq \hat g_{\lambda}(u)$. 
Hence, $g(u) \leq \hat g_{\lambda}(u)$ for all $u \in \reals$ and $\lambda > 0$ 
(Fig. \ref{f-hinge-restrict}). 
Evaluating at $u_i = y_i - b$ and summing over all voxels $i \in \mathcal{V}_s$, 
we obtain
\[
v_s(y,b) = \sum_{i \in \mathcal{V}_s} g(y_i - b) 
\leq 
\sum_{i \in \mathcal{V}_s} \hat g_{\lambda}(y_i - b) = \hat v_s(y,b;\lambda).
\]
An upper bound on the restricted value count at a given point thus ensures the 
exact value count is bounded above as well. % This convex approximation arises commonly in stochastic programming and is analogous to supplanting VaR with CVaR \cite[\S 2]{NemirovskiShapiro:2006}.

We can guarantee our dose-volume constraints hold by enforcing specific limits 
on $\hat v$. 
If $\hat v_s(y,u;\lambda) \leq \phi_s(p)$, then $v_s(y,u) \leq \phi_s(p)$, and 
the upper dose-volume constraint, $D_s(p,y) \leq u$, is satisfied. 
Similarly, $\hat v_s(-y,-\ell;\lambda) \leq \phi_s(100-p)$ implies that 
$D_s(p,y) \leq \ell$. 
To simplify notation, we rewrite $\hat v_s(y,b;\lambda) \leq \phi$ as
\[
\sum_{i \in \mathcal{V}_s} (1 + \lambda(y_i-b))_+ \leq \phi.
\]
Since $\lambda > 0$, we can divide both sides of the inequality by $\lambda$. 
Letting $\alpha := \frac{1}{\lambda}$ and gathering all terms on the left-hand 
side, we obtain the inequality
\[
\sum_{i \in \mathcal{V}_s} (\alpha + (y_i - b))_+ - \alpha \phi \leq 0.
\]
The left-hand side of this inequality is a sum of convex functions of 
$(\alpha, y)$, and hence convex.
Note that while $\lambda$ was a parameter of our restricted value count 
function, $\alpha > 0$ can be an optimization variable, since the left-hand term 
is jointly convex in $\alpha$ and $y$. 
Additionally, we can replace the constraint $\alpha > 0$ with $\alpha \geq 0$ because 
when $\alpha = 0$, the constraint simplifies to $(y_i-b)_+ \leq 0$, which is 
equivalent to $y_i \leq b$ for all $i \in \mathcal{V}_s$. 
Certainly in this case, the condition $D_s(p,y) \leq b$ holds. 
Thus, by defining the functions
\[
\hat D_s^+(p,y,b,\alpha) 
	= \sum_{i \in \mathcal{V}_s} (\alpha + (y_i - b))_+ - \alpha \phi_s(p)
\]
for upper constraints and
\[
\hat D_s^-(p,y,b,\alpha) 
	= \sum_{i \in \mathcal{V}_s} (\alpha - (y_i - b))_+ - \alpha \phi_s(100-p),
\]
for lower constraints, each convex restriction can be represented by 
inequalities in terms of these functions: for $\alpha \geq 0$, $\hat D_s^+(p,y,u,\alpha) \leq 0$ 
implies $D_s(p,y) \leq u$, and $\hat D_s^-(p,y,\ell,\alpha) \leq 0$ implies 
$D_s(p,y) \geq \ell$.
Our convex formulation with restricted dose-volume constraints is
\BEQ\label{cvx-restrict}
\begin{array}{lll}
	\mbox{minimize} & f(y) \\
	\mbox{subject to} 
		& y=Ax \\
		& x \succeq 0, \quad \alpha \succeq 0 & \\
		& \hat D_s^+\left(p_{s,k},y,u_{s,k},\alpha_{s,k}^{(u)}\right) 
			\leq 0, & k=1,\ldots,K_s^{(u)}, \quad s=1,\ldots,S \\
		& \hat D_s^-\left(p_{s,k},y,\ell_{s,k},\alpha_{s,k}^{(\ell)}\right) 
			\leq 0, & k=1,\ldots,K_s^{(\ell)}, \quad s=1,\ldots,S \\
		& \ell_s^{\mathrm{avg}} \leq D_s^{\mathrm{avg}}(y) 
			\leq u_s^{\mathrm{avg}}, & s=1,\ldots,S \\
		& D_s^{\min}(y) \geq \ell_s^{\min}, & s=1,\ldots,S \\
		& D_s^{\max}(y) \leq u_s^{\max}, & s=1,\ldots,S,
\end{array}
\EEQ
where for every structure $s$, we index the parameters of its upper dose-volume 
constraints with $k=1,\ldots,K_s^{(u)}$, and its lower dose-volume 
constraints with $k=1,\ldots,K_s^{(\ell)}$. 
We include a separate optimization variable, $\alpha_{s,k}$, in each dose-volume constraint to represent the inverse slope of its convex restriction and stack 
these variables in a vector 
$\alpha := (\alpha^{(\ell)}, \alpha^{(u)})$. 
Optimizing over $\alpha$ in addition to $(x,y)$ ensures we obtain the best 
hinge loss approximation to the value count function. 
The above formulation is a restriction of our original problem: if 
$(x,y,\alpha)$ is feasible for Prob. \ref{cvx-restrict}, then $(x,y)$ is feasible 
for Prob. \ref{cvx-exact}.

\section{Refinements}\label{refinements}

\subsection{Two-Pass Refinement}

A solution $(x^*,y^*,\alpha^*)$ to Prob. \ref{cvx-restrict} satisfies our 
restricted dose-volume constraints, so it is feasible for our original 
Prob. \ref{cvx-exact} with exact dose-volume constraints. 
However, since the convex restriction enforces an upper bound on the 
restricted value count function $\hat v$, the feasible set of Prob. \ref{cvx-restrict} 
is a subset of the feasible set of Prob. \ref{cvx-exact}, and $(x^*, y^*)$ 
may not be optimal for the latter. 
One way to improve our solution is to bound only the minimum number of voxels 
in each structure required to satisfy the dose-volume constraint. 
A good heuristic is to select those voxels $i$ that receive a dose $y_i^*$, 
which satisfies the associated dose-volume bound by the largest margin, and 
re-solve the problem with the convex restriction replaced by bounds on just 
these voxels. 
The solution of this second pass, $(x^{**},y^{**})$, will achieve an objective 
value $f(y^{**}) \leq f(y^*)$ while still satisfying our exact dose-volume 
constraints.

To make this precise, consider the lower dose-volume constraint 
$D_s(p,y) \geq \ell$. 
This is equivalent to $y_i \geq \ell$ for at least $\phi_s(p)$ voxels in 
structure $s$. 
Given $y^*$ from our first pass optimization, we compute the margin 
$\xi_i^* = (y_i^* - \ell)$ and select the $q_s = \lceil \phi_s(p) \rceil$ voxels 
$i \in \mathcal{V}_s$ with the largest values of $\xi_i^*$. 
Call this subset $\mathcal{Q}_s^- \subseteq \mathcal{V}_s$. 
Now, we replace $D_s(p,y) \geq \ell$ in Prob. \ref{cvx-exact} with the precise 
voxel constraints $y_i \geq \ell$ for all $i \in \mathcal{Q}_s^-$. 
On the second pass,
\[
v_s(y,\ell) 
= \sum_{i \in \mathcal{V}_s} \mathbbm{1}\{y_i \geq \ell\} 
\geq \sum_{i \in \mathcal{Q}_s^-} \mathbbm{1}\{y_i \geq \ell\} 
= q_s \geq \phi_s(p),
\]
so our upper dose-volume constraint is satisfied. 
An analogous argument with $q_s = \lceil \phi_s(100-p) \rceil$ and 
$\xi_i^* = (u - y_i^*)$ produces the subset $\mathcal{Q}_s^+$ for an upper 
dose-volume constraint $D_s(p,y) \leq u$. 
Given a solution $(x^*,y^*,\alpha^*)$ to Prob. \ref{cvx-restrict}, we repeat this 
process with every such constraint to obtain the second-pass problem formulation
\BEQ\label{cvx-2pass}
\begin{array}{lll}
	\mbox{minimize}  & f(y) \\
	\mbox{subject to} 
		& y=Ax & \\
		& x \succeq 0 & \\
		& y_i \leq u_{s,k} \; \forall i \in \mathcal{Q}_{s,k}^+,
			& k=1,\ldots,K_s^{(u)}, \quad s=1,\ldots,S \\
		& y_i \geq \ell_{s,k} \; \; \forall i \in \mathcal{Q}_{s,k}^-,
			& k=1,\ldots,K_s^{(\ell)}, \quad s=1,\ldots,S \\
		& \ell_s^{\mathrm{avg}} \leq D_s^{\mathrm{avg}}(y) 
			\leq u_s^{\mathrm{avg}}, & s=1,\ldots,S \\
		& D_s^{\min}(y) \geq \ell_s^{\min}, & s=1,\ldots,S \\
		& D_s^{\max}(y) \leq u_s^{\max}, & s=1,\ldots,S,
\end{array}
\EEQ
where the voxel subsets are indexed with $k=1,\ldots,K_s^{(u)}$ for upper 
dose-volume constraints, and $k=1\ldots,K_s^{(\ell)}$ for lower dose-volume 
constraints. 
We can warm start our solver at $(x^*, y^*)$ to speed up the second pass 
optimization.

\begin{algdesc}
	\label{2pass-algo}
	\emph{Two-pass algorithm.} 
	\begin{tabbing}
		{\bf given} a dose matrix $A\in \reals^{m\times n}$, a prescribed dose 
			vector $d \in \reals^m$, \\and a set of dose-volume constraints 
			$\mathcal{C}$. \\
		1.\ \emph{First pass.} Obtain the solution $(x^*, y^*, \alpha^*)$ to 
			Prob. \ref{cvx-restrict}.\\
		{\bf for each} $(\ell,p,s) \in \mathcal{C}$ {\bf do} \\
			\qquad \= 2a.\ \emph{Compute margins.} Calculate 
				$\xi_i^* = y_i^* - \ell$ for all $i \in \mathcal{V}_s$.\\
			\qquad \= 2b. \emph{Sort margins.} Sort 
				$\{\xi_i^*\}_{i \in \mathcal{V}_s}$ in ascending order to form a 
				set $\xi_s$. \\
			\qquad \= 2c.\ \emph{Identify voxel subset.} Select the 
				$\lceil \phi_s(p) \rceil$ largest values $\xi_i \in \xi_s$\\ 
			\> \qquad and include their indices $i$ in $\mathcal{Q}_{s,k}^-$.\\
		{\bf end for} \\
			{\bf for each} $(u,p,s) \in \mathcal{C}$ {\bf do} \\
			\qquad \= 3a.\ \emph{Compute margins.} Calculate 
				$\xi_i^* = u - y_i^*$ for all $i \in \mathcal{V}_s$.\\
			\qquad \= 3b. \emph{Sort margins.} Sort 
				$\{\xi_i^*\}_{i \in \mathcal{V}_s}$ in ascending order to form a 
				set $\xi_s$. \\
			\qquad \= 3c.\ \emph{Identify voxel subset.} Select the 
				$\lceil \phi_s(100-p) \rceil$ largest values $\xi_i \in \xi_s$ \\ 
			\> \qquad and include their indices $i$ in $\mathcal{Q}_{s,k}^+$.\\
			{\bf end for} \\
		4.\ \emph{Second pass.} Obtain the solution $(x^{**},y^{**})$ to 
			Prob. \ref{cvx-2pass} using $(x^*, y^*)$ as a warm start point.
	\end{tabbing}
\end{algdesc}

\subsection{Dose Constraints with Slack}

If our dose constraints are too strict, Prob. \ref{cvx-restrict} may not 
have a solution. 
This can arise even if the feasible set for our original 
Prob. \ref{cvx-exact} is non-empty, since our convex restriction enforces 
more stringent bounds on the dose distribution. 
To ensure the first pass of our algorithm always supplies a solution, we 
introduce a slack variable $\delta \in \reals_+$ to the bounds of each dose 
constraint, mapping lower bounds $\ell \mapsto (\ell - \delta)$ and upper bounds 
$u \mapsto (u + \delta)$. 
This creates soft constraints that need not be met precisely by the solution. 
Our problem reformulated with restricted dose-volume constraints and slack is 
\BEQ\label{cvx-restrict-slack}
\begin{array}{lll}
	\mbox{minimize}  & f(y) & \\
	\mbox{subject to} 
		& y=Ax & \\
		& x \succeq 0, \quad \alpha \succeq 0, \quad \delta \succeq 0 & \\
		& \hat D_s^+\left(p_{s,k},y,u_{s,k} + \delta_{s,k}^{(u)},
			\alpha_{s,k}^{(u)}\right) \leq 0, 
			& k=1,\ldots,K_s^{(u)}, \quad s=1,\ldots,S \\
		& \hat D_s^-\left(p_{s,k},y,\ell_{s,k} - \delta_{s,k}^{(\ell)},
			\alpha_{s,k}^{(\ell)}\right) \leq 0, 
			& k=1,\ldots,K_s^{(\ell)}, \quad s=1,\ldots,S \\
		& \ell_s^{\mathrm{avg}}-\delta_s^{\mathrm{avg},(\ell)} 
			\leq D_s^{\mathrm{avg}}(y) 
			\leq u_s^{\mathrm{avg}}+\delta_s^{\mathrm{avg},(u)}, 
			& s=1,\ldots,S \\
		& D_s^{\min}(y) \geq \ell_s^{\min}-\delta_s^{\min}, & s=1,\ldots,S \\
		& D_s^{\max}(y) \leq u_s^{\max}+\delta_s^{\max}, & s=1,\ldots,S,
\end{array}
\EEQ
Note that $\delta := (\delta^{(u)}, \delta^{(\ell)})$ is a variable in the optimization, and the value of each 
$\delta_{s,k}$ indicates the amount (in units of delivered dose, \eg, Gy) by 
which each bound is weakened in the solution.

We can incorporate soft constraints into the two-pass algorithm as well. 
On the first pass, we solve Prob. \ref{cvx-restrict-slack} to obtain the
optimal variables $(x^*, y^*, \alpha^*)$ and the optimal slacks $\delta^*$. 
Our margin for selecting $\mathcal{Q}_s$ is now computed with respect to the 
slack bound, \ie, $\xi_i^* = (y_i^* - \ell + \delta^*)$ for lower-volume dose 
constraints, and $\xi_i^* = (u + \delta^* - y_i^*)$ for upper dose-volume 
constraints. 
Finally, we weaken the bounds in Prob. \ref{cvx-2pass} by $\delta^*$, giving 
us the reformulated second pass optimization with slack dose-volume constraints
\BEQ\label{cvx-2pass-slack}
\begin{array}{lll}
	\mbox{minimize} & f(y)  \\
	\mbox{subject to} 
		& y=Ax  \\
		& x \succeq 0 \\
		& y_i \leq u_{s,k}+\delta_{s,k}^{(u)*} \; 
			\forall i \in \mathcal{Q}_{s,k}^+, 
			& k=1,\ldots,K_s^{(u)}, \quad s=1,\ldots,S \\
		& y_i \geq \ell_{s,k}-\delta_{s,k}^{(\ell)*} \; \; 
			\forall i \in \mathcal{Q}_{s,k}^-, 
			& k=1,\ldots,K_s^{(\ell)}, \quad s=1,\ldots,S \\
		& \ell_s^{\mathrm{avg}}-\delta_s^{\mathrm{avg},(\ell)*} 
			\leq D_s^{\mathrm{avg}}(y) 
			\leq u_s^{\mathrm{avg}}+\delta_s^{\mathrm{avg},(u)*}, 
			& s=1,\ldots,S \\
		& D_s^{\min}(y) \geq \ell_s^{\min} - \delta_s^{\min*}, & s=1,\ldots,S \\
		& D_s^{\max}(y) \leq u_s^{\max} + \delta_s^{\max*}, & s=1,\ldots,S.
\end{array}
\EEQ

\begin{algdesc}
	\label{2pass-slack-algo}
	\emph{Two-pass algorithm with slack.} 
	\begin{tabbing}
		{\bf given} a dose matrix $A\in \reals^{m\times n}$, 
			a prescribed dose vector $d \in \reals^m$, \\
			and a set of dose-volume constraints $\mathcal{C}$. \\
		1.\ \emph{First pass.} Obtain the solution $(x^*,y^*,\alpha^*, \delta^*)$ 
			to Prob. \ref{cvx-restrict-slack}.\\
		{\bf for each} $(\delta^*,\ell,p,s) \in \mathcal{C}$ {\bf do} \\
		\qquad \= 2a.\ \emph{Compute margins.} Calculate 
			$\xi_i^* = y_i^* - \ell + \delta^*$ for all $i \in \mathcal{V}_s$.\\
		\qquad \= 2b. \emph{Sort margins.} Sort 
			$\{\xi_i^*\}_{i \in \mathcal{V}_s}$ in ascending order to form a 
			set $\xi_s$. \\
		\qquad \= 2c.\ \emph{Identify voxel subset.} Select the 
			$\lceil \phi_s(p) \rceil$ largest values $\xi_i \in \xi_s$ \\ 
		\> \qquad and include their indices $i$ in $\mathcal{Q}_{s,k}^-$.\\
		{\bf end for} \\
		{\bf for each} $(\delta^*,u,p,s) \in \mathcal{C}$ {\bf do} \\
		\qquad \= 3a.\ \emph{Compute margins.} Calculate 
			$\xi_i^* = u + \delta^* - y_i^*$ for all $i \in \mathcal{V}_s$.\\
		\qquad \= 3b. \emph{Sort margins.} Sort 
			$\{\xi_i^*\}_{i \in \mathcal{V}_s}$ in ascending order to form a set 
			$\xi_s$. \\
		\qquad \= 3c.\ \emph{Identify voxel subset.} Select the 
			$\lceil \phi_s(100 - p) \rceil$ largest values $\xi_i \in \xi_s$ \\ 
		\> \qquad and include their indices $i$ in $\mathcal{Q}_{s,k}^+$.\\
		{\bf end for} \\
		4.\ \emph{Second pass.} Obtain the solution $(x^{**}, y^{**})$ to 
			Prob. \ref{cvx-2pass-slack} using $(x^*, y^*)$ as a warm start
			point.
	\end{tabbing}
\end{algdesc}

\section{Implementation}\label{conrad}

We implement our radiation treatment planning methodology with \conrad{}, a 
Python-embedded open-source software package based on the convex programming 
library, CVXPY \cite{DiamondBoyd:2016}, using the convex solvers SCS \cite{OCPBoyd:2016} 
and ECOS \cite{DCBoyd:2013}. 
\conrad{} provides a simple, intuitive interface for ingesting patient data, 
constructing plans based on a clinical prescription, and visualizing the 
dose-volume histograms of the result. 
It allows the user to add dose constraints using syntax familiar to clinicians. 
Since \conrad{} is an ordinary Python library, it can be easily integrated into 
existing data processing pipelines.

The following code imports a prescription, solves for the optimal treatment 
plan without dose constraints, and plots the DVH curves for all the patient 
structures. 
The $m \times n$ dose-influence matrix \texttt{A} can be encoded as a NumPy 
\texttt{ndarray} or any of several sparse representations in Python. 
The $m$-length vector \texttt{voxel\_labels} enumerates the index of the assigned
structure for each voxel in the patient volume.
\begin{verbatim}
import conrad

# Construct the case with no dose constraints
case = conrad.Case()
case.prescription = "/Documents/prescriptions/rx_patient_01.yaml"
case.physics.dose_matrix = A
case.physics.voxel_labels = voxel_labels
graphics = conrad.CasePlotter(case)

# Solve with a single pass and no slack
status, run = case.plan(solver="ECOS", use_slack=False, use_2pass=False)

print("Problem feasible?:\n{}".format(status))
print("Dose summary:\n{}".format(case.dose_summary_string))

# Display color-coded plot of all DVH curves
graphics.plot(run, show=True)
\end{verbatim}

A \texttt{Case} object comprises \texttt{Anatomy}, \texttt{Physics}, 
\texttt{Prescription} and \texttt{PlanningProblem} objects. Prior to planning, the case's \texttt{Anatomy} and \texttt{Physics} objects must be 
built.
The user can either build the case's \texttt{Anatomy} by adding structures programmatically (with 
data on each structure's name, index, identity as target/non-target, and 
desired dose) or by ingesting a prescription, which can be supplied as a Python dictionary or as a YAML or JSON file formatted for 
\conrad{}'s parser. 
The minimum information required for the case's \texttt{Physics} object are the $m \times n$ dose matrix and a $m$-length vector of voxel labels.
The case's \texttt{Prescription} object can be used to populate the 
\texttt{Anatomy} object or to keep track of clinical guidelines and objectives. It can also be left empty. 

The case's \texttt{PlanningProblem} object builds and solves optimization 
problems based on the structures in the case anatomy and any constraints 
assigned to those structures. The \texttt{PlanningProblem} is not exposed to the 
user. Instead, users form a treatment plan by calling the case's \texttt{plan()} method,
which returns a \texttt{bool} status indicating whether the specified problem
was feasible, along with a \texttt{RunRecord} object that carries solver 
performance data, optimal variables, and DVH curves.

Before planning a case, the user can add, remove, or modify dose constraints to 
any structure. Thus, even when a case has an assigned prescription, the dose
constraints attached to each structure in the case anatomy may differ
from the constraints specified in the prescription. For example, the prescribed 
constraints may correspond to clinical guidelines, while the constraints used
during planning may be chosen arbitrarily by the user to obtain plans with 
desirable dose properties.

After planning a case, users can plot the DVH curves, retrieve and print summaries of dose statistics for each structure, and when applicable, display a report of whether the current plan satisfies each constraint listed in the prescription. A case can be 
re-planned with different objective weights or dose constraints on any 
structure. The \conrad{} library provides a \texttt{PlanningHistory} object
to retain and manage results from prior runs.

The following code adds a dose-volume constraint to the PTV from our previous 
case, allowing at most 20\% of the PTV's voxels to receive more than 70 Gy. 
Algorithm \ref{2pass-algo} is then applied to obtain an optimal beam output. 

\begin{verbatim}
# Constraint allows at most 20% of PTV voxels to receive dose above 70 Gy 
case.anatomy["PTV"].constraints += D(20) <= 70 * Gy

# Solve with two-pass algorithm and no slack
_, run = case.plan(solver="ECOS", use_slack=False, use_2pass=True)
print("x PASS 1: {}".format(run.x_pass1))
print("x PASS 2: {}".format(run.x_pass2))

# Plot DVH curves from first (dashed) and second pass (solid)
graphics.plot(run, show=False, ls="--")
graphics.plot(run, second_pass=True, show=True, clear=False, legend=True)
\end{verbatim}

%%%%%% BEGIN INACTIVE BLOCK. 
\if{}
\conrad{} also offers a graphical user interface for adding dose-volume constraints 
to the DVH curves, which provides clinicians a visual reference for setting 
bounds. TODO: REFERENCE A FIGURE WITH POINT AND CLICK INTERFACE
\fi
%%%%%% END INACTIVE BLOCK

\section{Examples}\label{examples}

\subsection{Basic Functionality}
We present results illustrating the methods described in this paper: 
approximating dose-volume constraints via convex restrictions, two-pass
refinement of plans with dose-volume constraints, and handling
incompatible constraints with slack variables.

\paragraph{Problem Instances.}
We demonstrate the basic functionality on a head-and-neck case
expressed as a VMAT aperture re-weighting problem.
The case contains 360 apertures in four arcs, 270,000 voxels distributed
across 17 planning structures, including the PTV treated to 66 Gy, two 
auxiliary targets treated to 60 Gy, several OARs, and generic body voxels. 

To test the handling of dose-volume constraints, we plan the case
with no dose constraints and then re-plan with a single dose-volume
constraint applied to the PTV, namely $D(20) \le 70$ Gy.
We run the two-pass algorithm and compare the plans obtained by 
applying restricted and exact versions of the aforementioned dose-volume
constraint. Finally, we test the slack method by planning the case with two incompatible
dose-volume constraints: $D(98) \ge 66$ Gy on the PTV and
$D(20) \le 20$ Gy on the spinal cord. 
We compare the results from enforcing the PTV constraint alone, both 
constraints without slack, and both constraints with slack allowed.

\paragraph{Computational Details.}
The size of the dose matrix passed from \conrad{} to the convex solvers in
the backend varied depending on the dose constraints.  
When no minimum, maximum, or dose-volume constraints were applied to a non-target structure, the submatrix for that structure was replaced
with a mean dose representation, thereby eliminating $|\mathcal{V}_s| - 1$ rows from the problem matrix. 
In particular, the matrix representing the full dose on the targets and
mean dose for non-targets had dimensions 11,141 $\times$ 360, and the 
matrix including the full dose on the spinal cord was 15,000 $\times$ 360.
The \conrad{} problem request was formulated as a convex program
in CVXPY and passed to a GPU-based implementation of the convex 
solver SCS.
Calculations were performed on a cluster with 32-core, 2.20 GHz Intel Xeon 
E5-4620 CPU and a nVidia TitanX graphics card.

\begin{figure}
	\begin{center}
		\begin{subfigure}[b]{0.48\textwidth}
			\caption{Unconstrained}
			\label{f-dvh-constraints-a}
			\includegraphics[width=\textwidth]{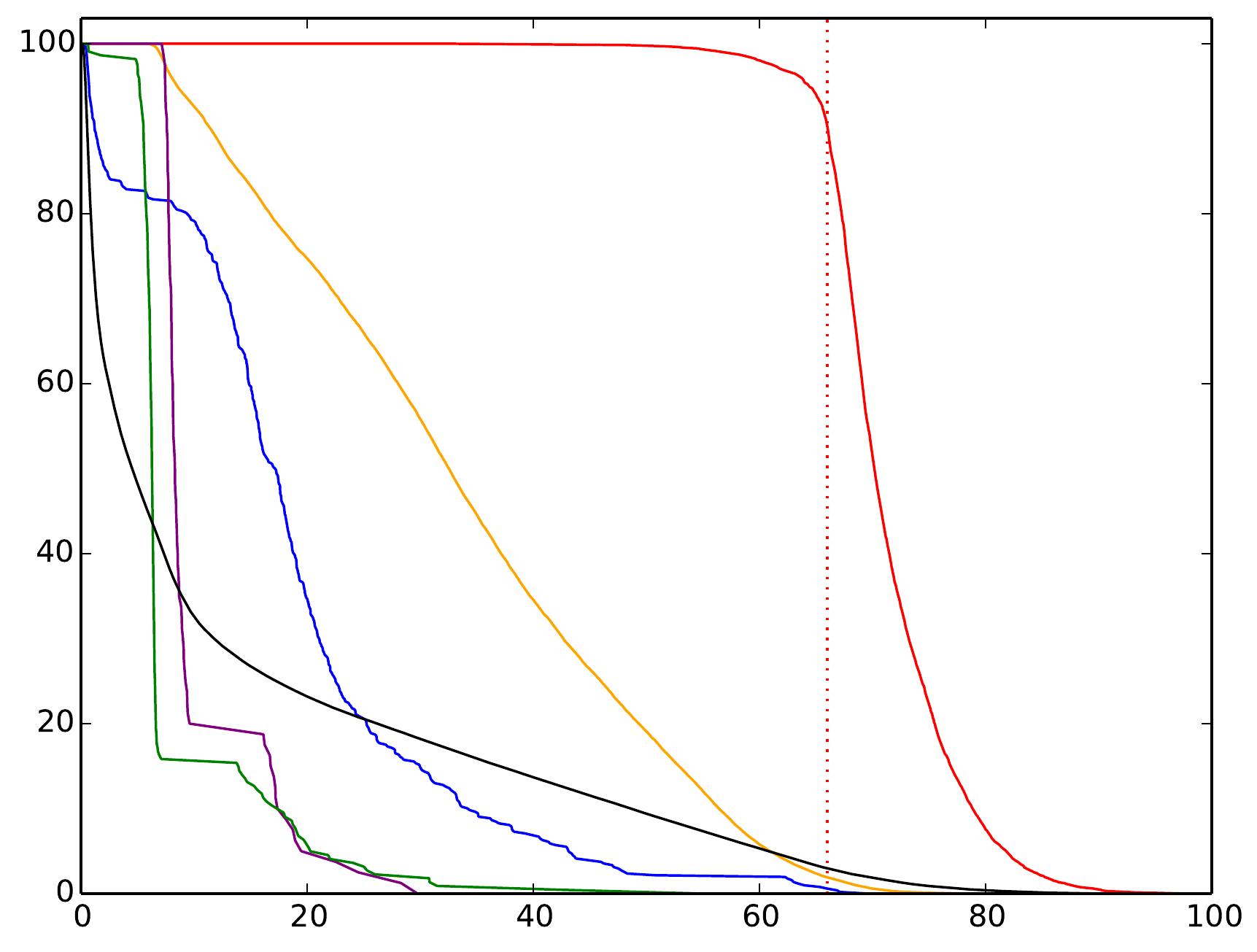}
		\end{subfigure}
		\begin{subfigure}[b]{0.48\textwidth}
			\caption{Add Dose-Volume Constraint}
			\label{f-dvh-constraints-b}
			\includegraphics[width=\textwidth]{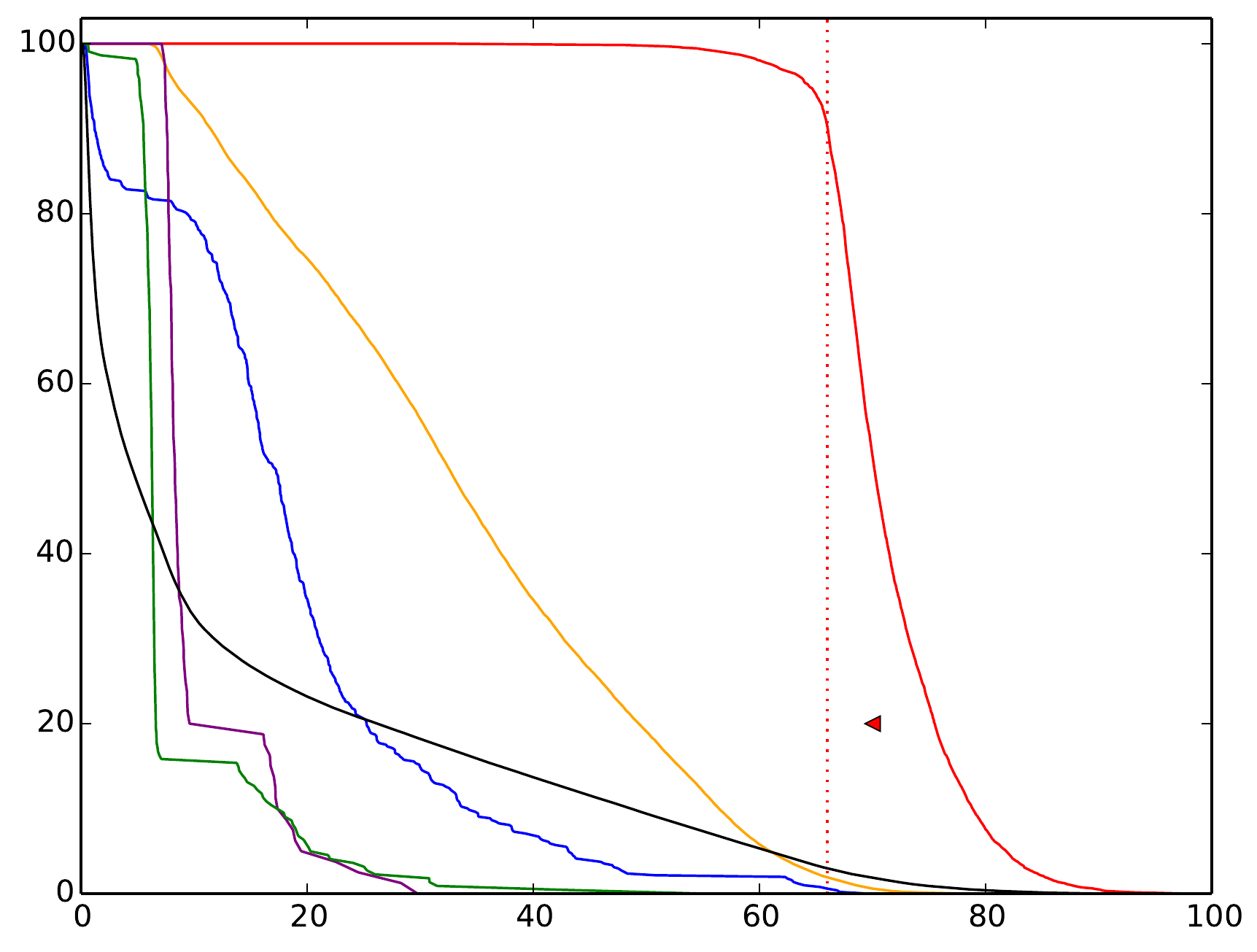}
		\end{subfigure}
		\begin{subfigure}[b]{0.48\textwidth}
			\caption{Constrained}
			\label{f-dvh-constraints-c}
			\includegraphics[width=\textwidth]{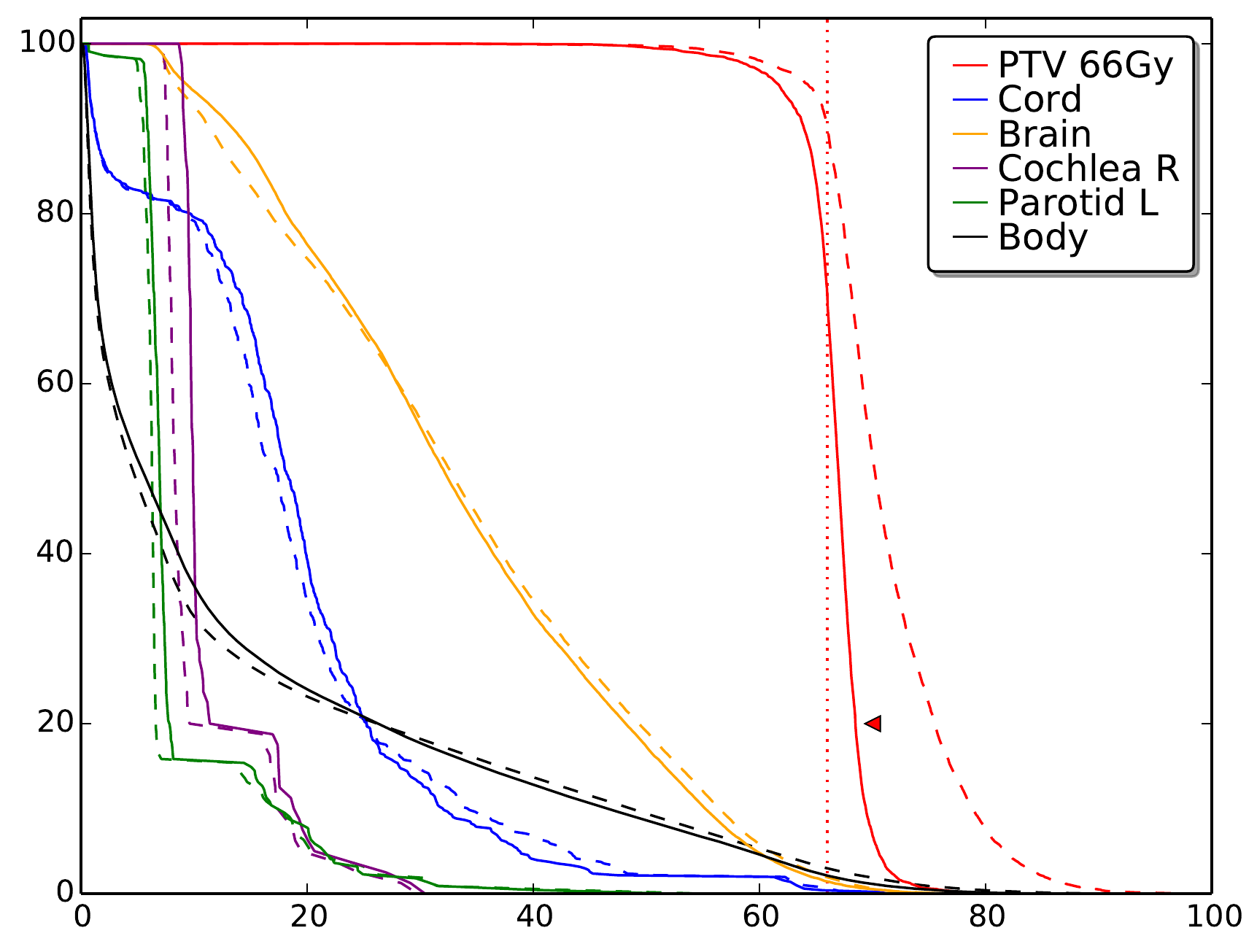}
		\end{subfigure}
		\caption{DVH curves for PTV and several OARs from the 4-arc VMAT 
		head-and-neck case. (a) Plan without any dose constraints. (b) Add a 
		dose-volume constraint $D(20) \leq 70$ Gy. (c) Re-plan with the new 
		constraint. The unconstrained plan is shown in dashed lines.}
		\label{f-dvh-constraints}
	\end{center}
\end{figure}

\paragraph{Clinical Results: Dose-Volume Constraints.}
Fig. \ref{f-dvh-constraints-a} depicts the DVH curves of the plan produced without any dose constraints. The PTV curve is shown in red with a dotted vertical line marking its prescribed dose of 66 Gy. The rest are DVHs for the OARs and generic body voxels. This solution to the unconstrained problem already gives a fairly good treatment plan. The DVH of the right cochlea and left parotid are pushed far left, so only 15--20\% of their volume exceed 10 Gy, and almost no voxels are dosed above 30 Gy. The spinal cord receives somewhat more radiation, while the worst case is the brain with a slow, nearly linear drop-off to about 75 Gy.

The PTV curve begins to fall at 66 Gy, but does not reach zero until nearly 95 Gy. To reduce this overdosing, we add a dose-volume constraint that limits no more than 20\% of the PTV to receive over 70 Gy, as indicated by the red, left-pointing arrow in Fig. \ref{f-dvh-constraints-b}, and re-plan the case. The resulting DVH curves are depicted as solid lines in Fig. \ref{f-dvh-constraints-c}, while the dashed curves represent the original plan. Under the new plan, the PTV curve has been pushed left at the arrow, and its drop-off around 66 Gy is steeper, meaning all voxel doses are closer to the prescription. Moreover, our OARs are minimally affected. We have reduced overdosing to the PTV without significantly increasing radiation to other organs.

\begin{figure}
	\begin{center}
		\begin{subfigure}[b]{0.48\textwidth}
			\caption{First Pass}
			\label{f-2pass-a}
			\includegraphics[width=\textwidth]{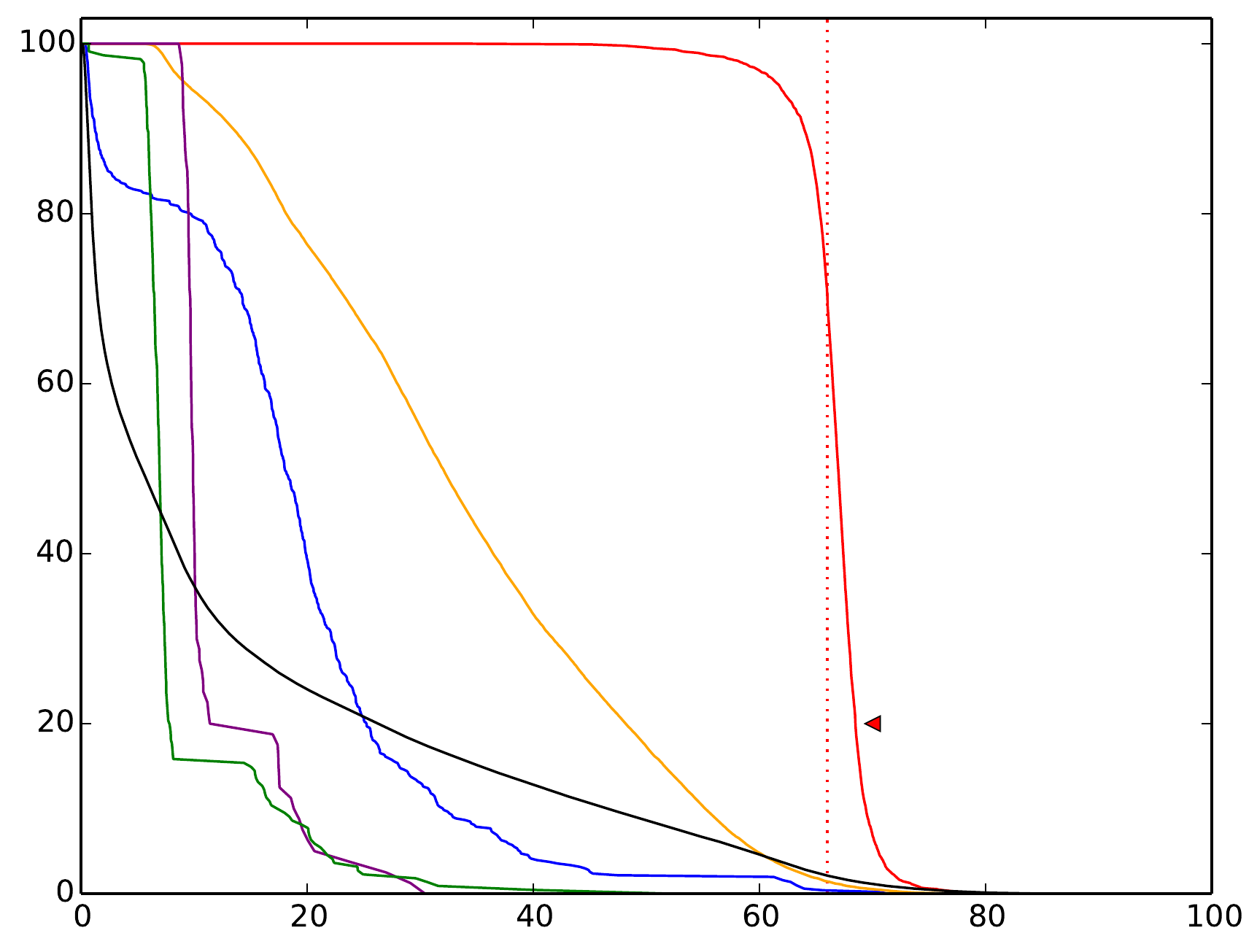}
		\end{subfigure}
		\begin{subfigure}[b]{0.48\textwidth}
			\caption{Second Pass}
			\label{f-2pass-b}
			\includegraphics[width=\textwidth]{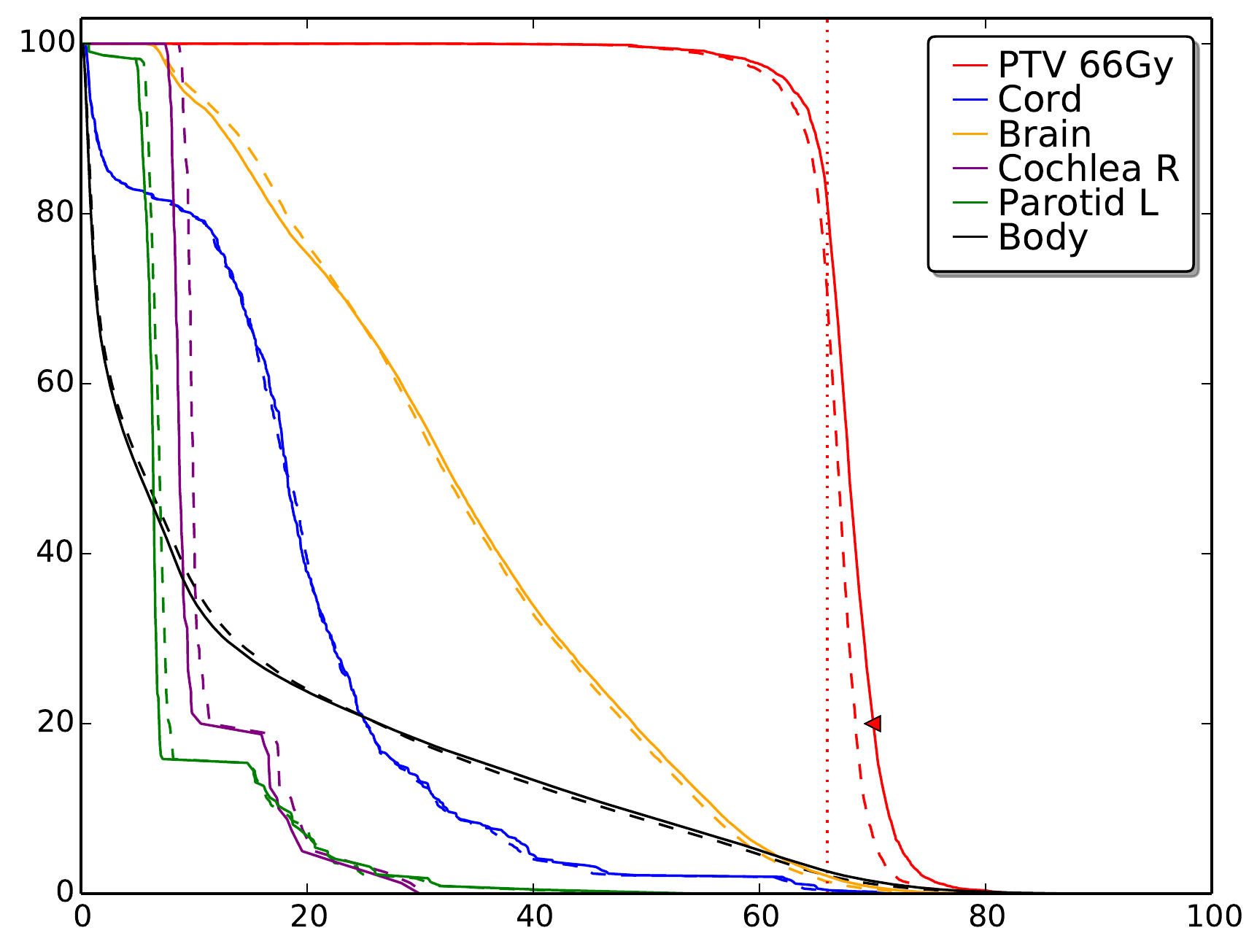}
		\end{subfigure}
		\caption{DVH curves for a two-pass algorithm with a single dose-volume 
		constraint on the PTV. (a) On the first pass, the constraint is met with 
		a margin of about 0.5 Gy. (b) On the second pass, the constraint is met 
		tightly with small gains elsewhere.}
		\label{f-2pass}
	\end{center}
\end{figure}

\paragraph{Clinical Results: Two-Pass Algorithm.}
A close inspection of Fig. \ref{f-dvh-constraints-c} reveals a gap between the PTV curve and the DVH constraint arrow. This is due to the conservative nature of the convex restriction, which overestimates the number of voxel violations. We can eliminate this gap and improve our overall objective with the two-pass algorithm. Fig. \ref{f-2pass-a} shows the plan from the first pass. There is a margin of about 0.5 Gy between the PTV curve and arrow, meaning at most 20\% of the PTV receives over 69.5 Gy, a more restrictive solution than required by our constraint $D(20) \leq 70$ Gy.

This margin disappears in Fig. \ref{f-2pass-b}. Here, the dashed lines depict the first pass solution, and the solid lines come from the second pass. The second pass PTV curve falls precisely on the center of the left-facing arrow, meaning the dose-volume constraint is tight. In addition, the DVH curves for the right cochlea and left parotid have shifted leftward, indicating they now receive less radiation. By replacing our convex restriction with exact voxel constraints, we are able to make gains in our OAR clinical objectives while still fulfilling the DVH constraint. 

\paragraph{Clinical Results: Constraints with Slack.}
So far, we have specified only one dose constraint. A problem may become infeasible when multiple constraints are enforced, either because the user-supplied bounds are too extreme or the convex restriction too severe in its overestimation of the voxel count. In such cases, we can still produce a plan that approximately conforms to the desired specifications by enabling slack constraints.

\begin{figure}
	\begin{center}
		\begin{subfigure}[b]{0.48\textwidth}
			\caption{Single Constraint, No Slack}
			\label{f-slack-a}
			\includegraphics[width=\textwidth]{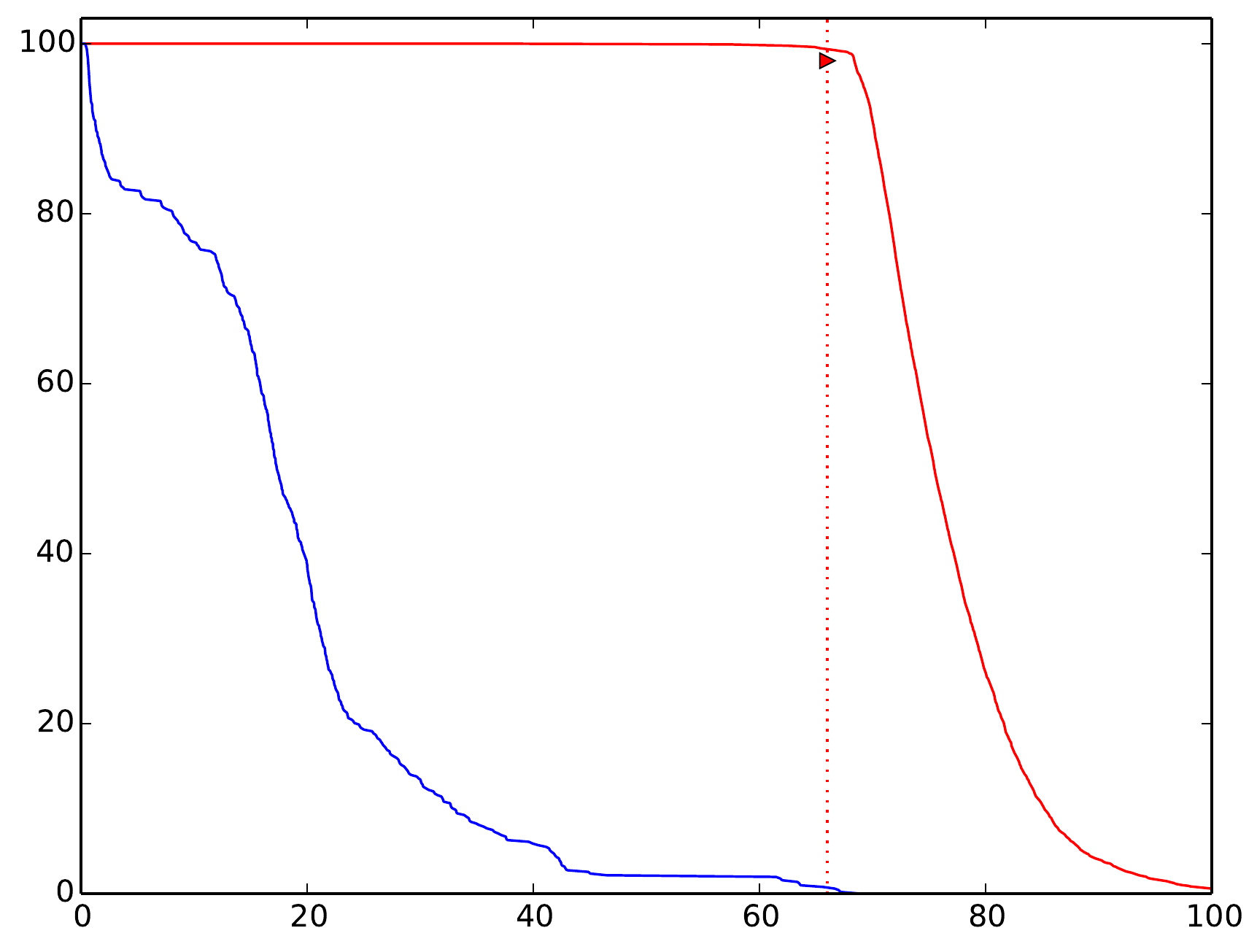}
		\end{subfigure}
		\begin{subfigure}[b]{0.48\textwidth}
			\caption{Two Constraints, No Slack}
			\label{f-slack-b}
			\includegraphics[width=\textwidth]{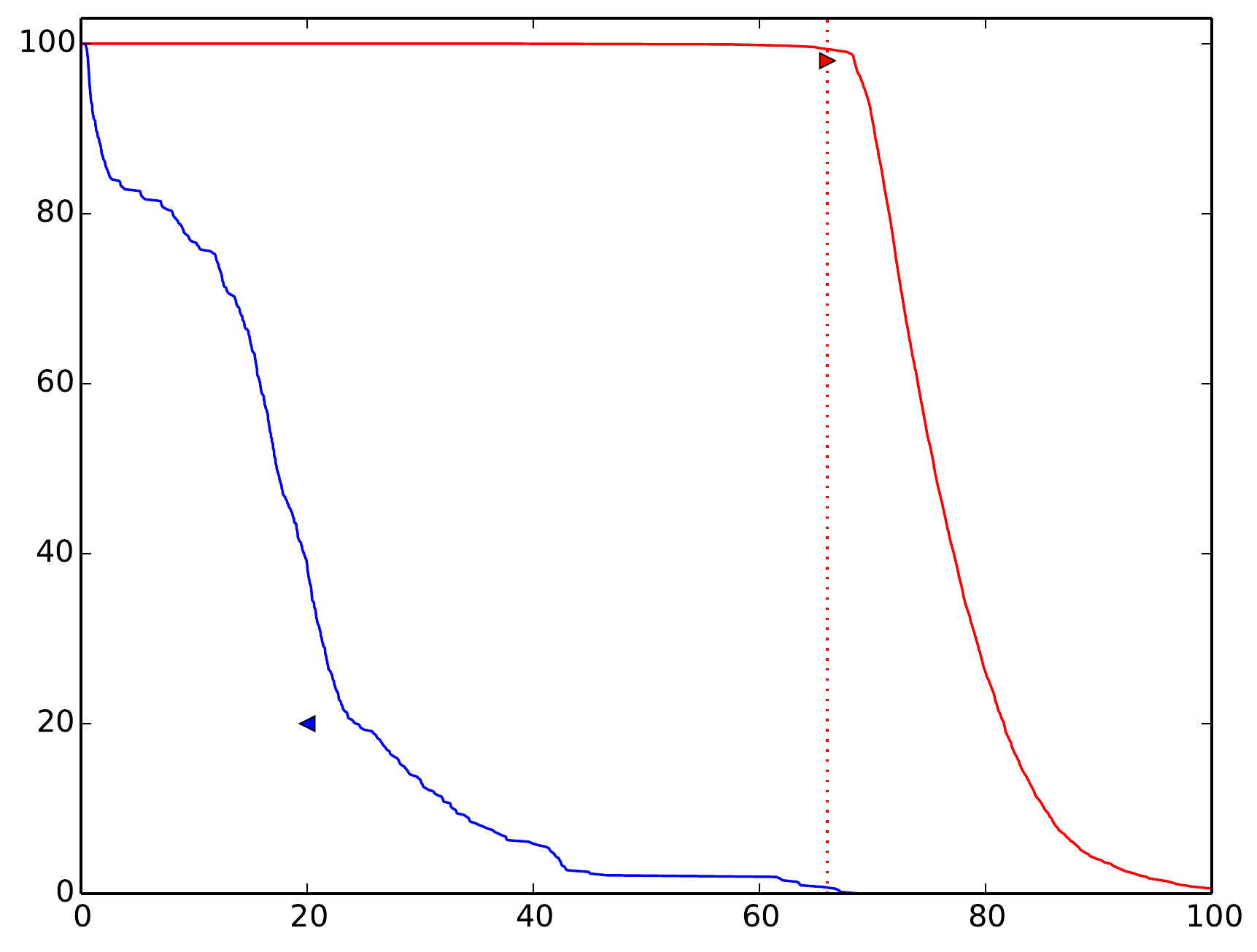}
		\end{subfigure}
		\begin{subfigure}[b]{0.48\textwidth}
			\caption{Two Constraints, Slack Allowed}
			\label{f-slack-c}
			\includegraphics[width=\textwidth]{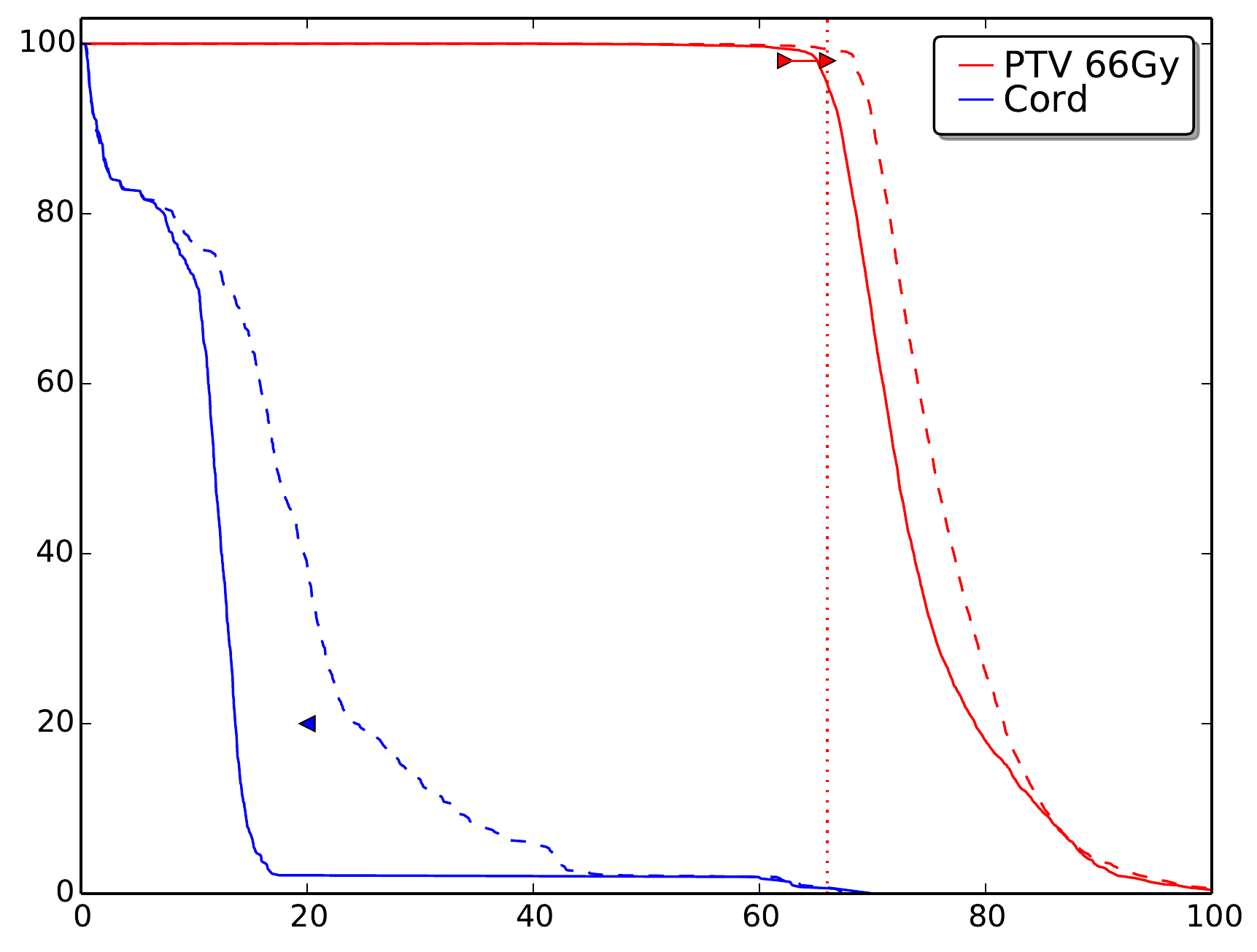}
		\end{subfigure}
		\caption{DVH curves for the PTV and spinal cord. (a) Plan without slack, 
		constraining $D(98) \geq 66$ Gy in the PTV. (b) A constraint 
		$D(20) \leq 20$ Gy is added to the spinal cord, rendering the problem 
		infeasible. (c) Re-plan with slack allowed. The spinal cord constraint 
		is met, but the PTV constraint relaxes by about 3 Gy.}
		\label{f-slack}
	\end{center}
\end{figure}

Fig. \ref{f-slack-a} depicts a plan created without slack. The single PTV constraint, $D(98) \geq 66$ Gy, is met with margin. However, when we add the OAR constraint $D(20) \leq 20$ Gy, as symbolized by the blue arrow in Fig. \ref{f-slack-b}, and re-plan the case, the optimizer tells us that the problem is infeasible. It is impossible to meet both (convex restricted) dose-volume constraints exactly. We thus re-plan allowing for slack bounds on these constraints. The resulting DVH curves are plotted in Fig. \ref{f-slack-c} with dotted lines representing the original plan and solid lines for the new plan. The PTV constraint has been relaxed by about 3 Gy, as symbolized by the red arrow shifting left behind the solid red curve to $(98, 63)$. This small concession allows us to satisfy the OAR constraint by a wide margin.

\subsection{Problem Scaling}

\paragraph{Problem Instances.}
We assess the performance of our algorithm on a larger prostate FMO problem. This case contains 74,453 voxels $\times$ 34,848 beamlets, encompassing a single PTV treated to 75.6 Gy, five OARs with various dose constraints, and generic body voxels. Approximately 226 million (roughly 10.6\%) of the entries in the dose matrix are non-zero. In our experiments, we used only a subset of 10,000 beamlets from this matrix.

We plan the case with the prescription detailed in Table \ref{t-prostate-rx}, which is adapted from the QUANTEC guidelines \cite{MarksYorke:2010}. The computational details are the same as in the head-and-neck case. As before, we analyze the results from a single pass and two-pass algorithm with and without slack allowed. We then re-plan the case with only the PTV dose constraint and compare its runtime and OAR overdose to the plans produced from the full prescription.

\begin{table}
	\caption{Prostate FMO Prescription
		\label{t-prostate-rx}}
	{\begin{tabular*}{\hsize}{@{\extracolsep{\fill}}p{1.25in}p{0.75in}p{0.75in}p{1.25in}@{}}
			\hline
			Structure & Target? & Dose (Gy) & Constraints (Gy)\\
			\hline
			Prostate & Yes & 75.6 & $D^{\mathrm{avg}} \geq 75.6$\\
			\hline
			Urethra & No & 0 & $D^{\mathrm{avg}} < 52.5$\\
			\hline
			Bladder & No & 0 & $D(85) < 80$ \newline
			$D(75) < 75$ \newline
			$D(65) < 70$ \newline
			$D(50) < 65$ \\
			\hline
			Rectum & No & 0 & $D(90) < 75$ \newline
			$D(85) < 70$ \newline 
			$D(50) < 65$ \\
			\hline
			L. Femoral Head & No & 0 & $D(95) < 50$ \\
			\hline
			R. Femoral Head & No & 0 & $D(95) < 50$ \\
			\hline
			Body & No & 0 & $D^{\mathrm{avg}} < 52.5$ \\
			\hline
	\end{tabular*}}
\end{table}

\paragraph{Timing Results.}
Our algorithm produces a plan that satisfies all dose constraints using a single pass with slack enabled. The optimization finishes in 426.9 seconds, about 2.5x the runtime of the head-and-neck case. A second pass takes approximately the same amount of time and does not result in significantly larger constraint margins. The presence or absence of slack also has little impact on the runtime, which varies by at most 7 seconds.

If we drop all except the mean dose constraints, the problem collapses into a linear program, greatly decreasing the runtime. The size of this reduction depends on the characteristics of the affected structures and the dose influence matrix. For example, in the head-and-neck case, the runtime falls by 87\% to a mere 24 seconds. Conversely, when adding dose-volume constraints, the initial constraint on a structure will have a greater impact on runtime than subsequent additions.

%%%%%% BEGIN INACTIVE BLOCK.
\if{}
Our algorithm produces a plan that satisfies all dose constraints using a single pass with slack enabled. The optimization finishes in 426.9 seconds, about 2.5x the runtime of the head-and-neck case. Table \ref{t-timing-scs-gpu} gives a comparison of the two cases. A second pass takes approximately the same amount of time and does not result in significantly larger constraint margins. The presence or absence of slack also has little impact on the runtime, which varies by at most 7 seconds.

If we drop all except the mean dose constraints, the problem collapses into a linear program, greatly decreasing the runtime: in the head-and-neck case, we see an 87\% reduction to 24 seconds, while the larger prostate problem speeds up by TODO. The size of the reduction depends on the characteristics of the affected structures and the dose influence matrix. Conversely, when adding dose-volume constraints, the initial constraint on a structure will have a greater impact on runtime than subsequent additions.

\begin{table}
	\begin{center}
		\caption{Timing Results with SCS, GPU
			\label{t-timing-scs-gpu}}
		{\begin{tabular}{|l||l|c|c|}
				\hline
				& & Head and Neck & Prostate\\
				\hline
				\multirow{4}{*}{Dimension} & Voxels & 270,000 & 74,453\\
				& Beams & 360 & 10,000\\
				& Structures & 17 & 7\\
				& Nonzero Entries & TODO & $226 \times 10^6$\\
				\hline
				\multirow{2}{*}{Runtime (sec)} & Mean Constraints & 24.0 & TODO \\
				& Full Prescription & 189.7 & 462.9\\
				\hline
		\end{tabular}}
	\end{center}
\end{table}
\fi
%%%%%% END INACTIVE BLOCK.

\section{Conclusion}\label{conclusion}
We have developed a convex formulation for the FMO problem that incorporates dose-volume constraints. Our model replaces each exact dose-volume constraint with a convex restriction, which overestimates the number of voxels that violate the clinician's desired threshold. This allows us to solve the problem quickly and efficiently using standard convex optimization algorithms. We also introduce two refinements: a two-pass algorithm and a model with slack. In the former, we improve our initial solution by re-optimizing with the restrictions replaced by bounds on a subset of voxels, enabling us to achieve a better objective that still satisfies the dose-volume constraints. The latter allows for soft bounds and is useful if the restricted constraints render the problem infeasible. We demonstrate the efficacy of our method on a VMAT head-and-neck case and a prostate case. Our algorithm consistently produces good treatment plans that fulfill all dose constraints when feasible. In problems with infeasible constraints, we are able to generate plans that minimize the dose violation while taking into account clinical goals, allowing clinicians to easily visualize trade-offs and select the plan that is best for the patient.

A variety of extensions to our two-pass algorithm are possible. For instance, we could rewrite the original problem as a mixed-integer linear program and use the solution of the convex restriction to warm start a branch-and-bound solver. More broadly, we could apply this starting point to accelerate any number of iterative approaches in the literature. Dose-volume constraints are often assigned different priorities in practice, and our algorithm may be easily adapted to accommodate such user-defined preferences, either through new penalties, changes in the slack, or additional passes that impose the constraints in a lexicographic order. These hybrid methods, which combine convex approximations with non-convex solution methods, offer an important avenue for future research.

\section*{Acknowledgements}
We thank Michael Folkerts for providing the anonymized dataset for the head and neck
VMAT reweighting case, and Peng Dong for the anonymized dataset for the prostate IMRT
case. This research was supported by the Stanford Graduate Fellowship, Stanford Bio-X Bowes Fellowship, and NIH Grant 5R01CA176553.

\clearpage
\nocite{RTOG1016}
\nocite{RTOG1115}
\bibliography{conrad_refs}

\end{document}